# Uncovering shifts in the history of Physics education: a systematic, NLP-based, thematic analysis of articles from *The Physics Teacher* and *Physics Education* journals (1966 - 2019)

Martina Caramaschi[1], Tor Ole B. Odden[2]

[1]Department of Physics and Astronomy "A. Righi", University of Bologna, Bologna, Italy
[2]Center for Computing in Science Education, University of Oslo, 0316 Oslo, Norway

## Abstract

This study explores the thematic evolution of articles in *The Physics Teacher* and *Physics Education* journals, over a critical period in modern history, from the Cold War era to the pre-pandemic world (1966 - 2019). Using an NLP-based inductive topic modeling approach, we identify recurring themes that have shaped the physics education literature, including content-based topics, teaching methodologies, laboratory practices, curriculum development, and the influence of Physics Education Research (PER). Our findings reveal both overarching trends and distinct thematic preferences between the journals. *Physics Education* has historically emphasized curriculum structures, social aspects of education, and interdisciplinary connections, whereas *The Physics Teacher* has focused more on pedagogical strategies, demonstrations, and practical teaching tools. Over the past three decades, both journals have increasingly incorporated discussions on technology, computation, and PER-driven instructional practices. By tracing these developments over five decades, this study provides a broader perspective on how physics education has responded to changing educational priorities, technological advancements, and research developments.

## 1. Introduction

The field of physics education has a long and rich international history. In both Europe and the United States physics education in its current form has been established for well over half a century, dating back to efforts in the 1960s to create standardized curricula that could apply across universities and could help high schools understand how they should prepare their students for university studies [1, 2, 3].

In the United States, these efforts were spurred on by the Cold War, which motivated the US government to invest significant resources in science education in order to build a competitive scientific workforce [4]. During that same period, European countries also worked to improve their science education curricula as part of broader efforts to establish Europe as a technological and scientific power and as a critical component of Europe's technological development. This development was driven in part by Cold War dynamics, which reinforced the strength of European nations to maintain competitiveness with both the United States and the Soviet Union [5].

Around the same time, physics teaching journals were launched in both the United States and Europe. In the US, *The Physics Teacher* (TPT) published its first issue in 1963, going on to publish a diverse range of articles, from pedagogical contributions for the teaching of fundamental physics concepts to historical perspectives to explore experimental breakthroughs in physics. Many of TPT's articles aimed to make physics engaging, show its application in everyday life, or provide detailed descriptions of innovative experimental setups for classroom instruction (i.e., "*Apparatus for Teaching Physics*" series). Shortly after, in the United Kingdom, *Physics Education* (PE) published its first issue in 1966, with articles focusing on both conceptual and practical aspects of teaching physics. For example, articles explored effective teaching strategies for diverse physics topics such as Dynamics and Special Relativity, provided a guide to design effective physics experiments, and illustrated common misconceptions in physics, alongside articles more focused on interdisciplinary aspects of physics, and the historical and philosophical evolution of the discipline.

These journals are similar, since both made an early commitment to exploring both theoretical and applied aspects of physics education, serving as a resource for teachers and students alike. Their publication activity has continued uninterrupted to the present day. As of 2020, *The Physics Teacher* has published approximately 15,000 articles, and *Physics Education* has published around 7,000. This body of literature holds the potential to reveal many interesting aspects of how physics teaching and learning has evolved over the period of publication. However, the scale of the literature has made it intractable to do traditional, by-hand literature reviews.

If this literature could be analysed, it would help us understand the themes addressed by the two journals over time, highlighting shifts in interests and the types of materials they have provided to the community of teachers and researchers. Additionally, we could examine the differences between the two journals, which, despite covering similar topics, each have their own unique identity and focus. Analysing these aspects could offer a broad perspective on the evolving landscape of physics teaching, which, in various ways, has been reflected in the articles they have published. The selection of topics and articles has likely been influenced by topical events and prevailing interests in the field. Furthermore, given that the two journals focus on different geographical areas, that is the North American context for *The Physics Teacher* and the European one for *Physics Education*, such a study may also reveal differences shaped by their respective educational systems.

In this article, we conduct a thematic analysis of over 50 years' worth of articles published in these two journals using a technique from the field of natural language processing, Latent Dirichlet Allocation (LDA). In brief, we have integrated LDA into a qualitative thematic analysis to identify key themes in the articles in order to characterise thematic trends of the two journals over time. This analysis reveals topics addressed by the journals, mapping interests that were related to both the historical context and evolving needs of the field.

In what follows, we begin with the background for the study and an overview of the literature, where we situate this study within the history of physics education and introduce the two journals (*The Physics Teacher* and *Physics Education*) in section 2. Additionally, we introduce the LDA method and highlight its previous applications within the field of science and physics education research. In section 3, we outline the aims and research questions guiding our study. Section 4 contains a detailed description of our methodology, where we explain how we processed the data and performed the inductive thematic analysis. In Section 5, we present the results of our analysis, beginning with an overview of the topics discovered across both journals combined. We then explore distinct trends within each journal separately, revealing peculiar interests and educational contexts in North America and Europe. Finally, in Sections 6 and 7, we discuss the implications of our findings, highlighting the differences and similarities revealed between the two educational systems and providing conclusions on the evolving discourse in physics education over time.

## 2. Background and literature review

### 2. 1. Overview of the history of physics education

In this work, we aim to thematically investigate literature on physics education from the 1960s to the present. Thus, it is useful to contextualise this study with a discussion of significant events and important shifts in physics education over this period. We base this discussion on the work of Otero and Meltzer (2016) [4], which provides a historical description of developments in United States physics education since the 1800s, as well as a dedicated chapter in the book "*125 Years The Physical Society and The Institute of Physics*" which provides context the development of physics education in Europe [6].

Physics education has shifted through several different phases over its history. Within the United states, in the years following World War II the political and social importance of science and technology grew, particularly due to the Cold War. This led to substantial investment in science education. For example, the establishment of the National Science Foundation (NSF) in 1950 and the launch of Sputnik in 1957 triggered a rapid expansion of teacher training programs and the development of new high school physics curricula, such as the Physical Science Study Committee (PSSC) (1956) and, later, Project Physics (active from 1962) [7]. These programs aimed to foster a more conceptual and inquiry-based approach to learning.

During the 1960s and 1970s, efforts to improve physics education focused on teacher training and the creation of innovative teaching materials. The Commission on College Physics, founded in 1960 with support from the NSF, worked to enhance physics instruction at the university level and improve the preparation of high school teachers. In 1972, the National Research Council recommended inquiry-based physics courses for teacher training. However, many of these efforts remained limited to a relatively small portion of the educational system.

A significant driver of change emerged during the 1970s and 1980s, when Physics Education Research (PER) began to establish itself as an academic discipline. Scholars such as A. B. Arons and R. Karplus advocated for the use of inquiry-based teaching methods [8]. Meanwhile, researchers like L. C. McDermott at the University of Washington began developing educational programs based on systematic studies of students' learning difficulties [9]. A key moment came in the late 1980s with the development of the Force Concept Inventory (FCI) (1985) [10, 11], a diagnostic tool for assessing students' understanding of mechanics.

A defining feature of the 1990s was the rise of conceptual physics, which emphasized qualitative understanding over mathematical formalism. This contributed to a significant increase in high school physics enrollments [4]. At the same time, computational physics gained prominence, driven by advances in computer technology and the growing recognition of computation as a third pillar of scientific inquiry alongside theory and experiment [12]. Graphical modeling tools and simulation software became increasingly integrated into physics education, allowing students to visualize complex systems, explore dynamic interactions, and develop an intuitive grasp of physical principles. By shifting the focus from purely analytical methods to computational approaches, these tools expanded access to physics learning and reshaped the way students engaged with scientific problem-solving.

Returning to the role of PER, from 2000 onwards, it gained greater academic recognition. Studies such as that of R. R. Hake [13] demonstrated that active learning and research-based methodologies lead to significant improvements in student learning. As a result, many universities began implementing innovative teaching strategies, including the use of technology in laboratories and learning based on multiple representations. The impact on teaching was evident even before, from the 1990s onward, as the integration of PER findings into classroom practice led to the development of research-based curricula and interactive engagement techniques. This shift was also reflected in the dissemination of research literature, which, as McDermott and Redish noted in the 1999 on the *American Journal of Physics*, aimed to serve not only physics education researchers but also "*the much larger community of physics instructors whose primary interest is in using the results from research as a guide for improving instruction*" [14]. Today, physics education in the United States continues to evolve, balancing technological advancements, diverse student needs, and ongoing debates about the most effective ways to teach and learn the subject.

Within Europe, the 1950s was characterised by a period of significant stagnation for physics education [6]. Physics teaching commonly relied heavily on rote memorization, with students expected to memorize formulas, definitions, and experiment descriptions, and little focus on fostering critical thinking or inquiry. This traditional approach limited the scope of learning and created a disconnect between the rapidly evolving field of physics and the methods used in the classroom. Although there were some localized efforts to innovate, such as the introduction of new experimental apparatus through the Nuffield Physics Projects in the late 1950s, the overall impact of these initiatives was gradual and often impeded by the inertia of long-standing teaching practices.

However, the post-WWII period also marked the beginning of important shifts that would gradually transform European physics education. One key event was the founding of CERN in 1954, which was driven by the ambition to challenge American leadership in scientific research. CERN's establishment symbolized Europe's commitment to scientific innovation and international collaboration, laying the foundation for a more modern and dynamic approach to science. This development was not only crucial for advancing high-energy physics research but also influenced the educational landscape, signaling the need for an overhaul in how physics was taught in Europe.

In the 1960s, Physics education in Europe began to change. The exchange of knowledge with the United States, especially through models like the aforementioned Project Physics Course, played a significant role in introducing new teaching approaches. These American-led initiatives, combined with US-led science policy, heavily influenced European physics education, particularly in high-energy physics research. At the same time, European countries recognized the need to foster their own technological advancements, resulting in the launch of several initiatives designed to strengthen scientific knowledge across the continent. One such example was CERN's *Science for All* program [5], organized since the 1960s to engage non-academic staff in scientific discussions, reflecting a broader movement to disseminate scientific knowledge and build a stronger scientific workforce in Europe.

This decade, therefore, marked a turning point for physics education in Europe, as efforts to modernise the curriculum and teaching methods gained momentum. The Nuffield Physics Project, whose development was influenced by the Physical Science Study Committee (PSSC), gained more traction. It represented a significant step toward a more inquiry-based, student-centered approach to learning, emphasising the importance of developing students' critical thinking skills and moving away from rote memorization. In fact, both PSSC and Nuffield Project reforms placed greater emphasis on the process of science, encouraging students to think like scientists rather than merely recalling facts [15].

Between the 1970s and 1980s, European physics education, especially in the UK, saw significant reforms, largely driven by the Institute of Physics (IOP). The IOP helped establish several key committees, producing reports on issues, such as *The shortage of physics teachers*, *Resources for Teaching School Physics*, and *Interface between physics, mathematics, and engineering*. There was general support for mature students to earn a physics degree. Amidst these changes, the shortage of qualified teachers became a major concern, prompting governmental action. In 1988, the publication *Physics in Higher Education* led to the introduction of the four-year MPhys/MSci degrees, shifting focus towards professional training for physicists [6].

In the 1990s, the UK faced a significant decline in the number of students taking A-level physics, a trend that raised concerns about the future supply of skilled scientists and engineers. Recognizing the need to attract more young people to physics, the IOP launched the 16-19 Physics Initiative in 1997. This initiative aimed to make physics courses more engaging, modern, and relevant, beyond just revising syllabuses. This effort was seen as essential for ensuring a high-caliber workforce for the future [6].

Across these shared histories, we note several interesting trends. First, philosophy and history have played significantly different roles within physics education systems of the USA and Europe. As Bevilacqua and Giannetto describe [16], in Europe the postwar period was marked by a division between the humanities and the sciences. However, for physicists with a classical education acquired during high school, this cultural split was not entirely accepted or embraced. In some ways, it prompted the scientific community to engage in historical and philosophical debates, fostering a strong interest in foundational philosophical questions and a well-established tradition in the history of physics. This tradition reached its peak around 1950 but remained deeply influential in the following years.

In the United States, Kuhn's critique of textbook science sparked significant debate, but it largely remained confined to the historiography of science. This led US historians to distance themselves from the field of education, as reflected in the 1970 MIT meeting. In contrast, from the 1980s onward, Europe increasingly integrated history and philosophy into physics education, embracing a broader intellectual tradition [16].

Table 1. Summary of historical events and contexts that shaped physics education in the USA and Europe, based on Otero and Meltzer (2016) and Lewis (1999).

| 1950s | <ul><li>Post WWII</li><li>Cold War</li><li>Sputnik launch</li><li>CERN foundation (1954)</li><li>National Science Foundation</li><li>PSSC</li><li>Development of high school and university physics curricula</li><li>General stagnation for european physics education</li></ul> |
|---|---|
| 1960s - 1970s | <ul><li>Project Physics (1962)</li><li>Conceptual and inquiry-based approach to learning.</li><li>Focus on teacher training</li></ul> |

|  | ● Development of teaching material<br>● Science for All at CERN<br>● Diffusion of Nuffield Science (and Physics ) Project (1957) |
|---|---|
| 1980s | ● Establishment of Physics Education Research as academic discipline<br>● Studies on students' learning difficulties<br>● Force Concept Inventory (1985)<br>● Shortage of Physics teachers in Europe<br>● 4-years degrees in Physics |
| 1990s | ● Conceptual physics and qualitative understanding over formalism<br>● Rise of Computational physics<br>● Simulators included into physics education<br>● Integration of PER findings into classroom practice |
| 2000s - 2010s | ● Student-driven exploration<br>● Innovation in lab design: computational tools and interactive, technology-integrated physics labs<br>● Focus on students skills (e.g., troubleshooting skills)<br>● Results from research as a guide for improving instruction |

Second, both systems have seen major shifts in the role of laboratory instruction and instructional use of experiments, tools, and software. According to May (2023) [17], post-WWII, the physics education community critiqued traditional "cookbook" labs, which led to debates about better aligning labs with conceptual learning goals and experimental skills. This period saw the rise of more student-centered approaches, with innovations like "free" and guided inquiry labs, where students explored experiments with more autonomy. The American Association of Physics Teachers (AAPT) played a key role, releasing lab recommendations in 1957, promoting the shift away from rigid procedures to more exploratory learning.

In the 1970s and 1980s, technology integration, such as computers and simulations, enhanced data collection, analysis, and measurement uncertainty studies. Meanwhile, PER began assessing the effectiveness of these new methods [18, 19], shifting focus from traditional assessments like written exams to more practical, performance-based evaluations.

By the late 20th century, evidence-based lab reforms were introduced, emphasizing goals such as hands-on experimentation, conceptual understanding, and collaborative skills. The AAPT's 1998 report [20] outlined these as key goals for introductory labs, marking a shift from "how" labs were structured to "what" they aimed to achieve. This period also saw further innovation in lab design, with increasing use of computational tools and troubleshooting skills. The focus was increasingly on student-driven exploration and data analysis, laying the groundwork for the more interactive and technology-integrated physics labs of the 21st century.

This history provides a useful context for the present literary analysis. The literature selected for our analyses consists of a large collection of articles from two journals that share similar objectives, but were founded by different societies and serve different communities. Specifically, we examine publications from *The Physics Teacher* and *Physics Education* from their inception until the pandemic. Below, we summarise key characteristics of these journals, including their publication period, affiliated societies and article styles and target.

## 2. 2. Literature: *The Physics Teacher* and *Physics Education*

**The Physics Teacher**
Launched in 1963, *The Physics Teacher* is published by AIP Publishing on behalf of the *American Association of Physics Teachers* (AAPT), an organisation that has played a central role in advancing physics education since its founding in 1930.

Established by Homer L. Dodge, Paul E. Klopsteg, and William S. Webb, AAPT was created with the goal to improve physics teaching and the dissemination of physics knowledge. Over the decades, leading figures such as Melba Philips helped emphasise the idea that effective teaching is not only essential for scientific progress but also for a broader societal benefit. Through *The Physics Teacher*, AAPT provides educators with research-based insights, classroom strategies, and discussions on contemporary challenges in physics teaching.

The journal publishes short peer-reviewed articles, covering a wide range of topics as the teaching of introductory physics, contemporary and applied physics, and the history of the discipline. Its content and materials are designed for educators at all levels, including secondary schools, colleges, and universities. For this study, we had access to its complete corpus of publications from its inception to 2020.

**Physics Education**
*Physics Education* is an international journal dedicated to advancing the teaching and learning of physics. *Physics Education* is published by IOP Publishing on behalf of the Institute of Physics (IOP) and its first publication was in 1966. IOP is a UK-based professional organisation that promotes physics research, education, and outreach.

Unlike *The Physics Teacher*, which primarily serves the North American physics education community, *Physics Education* has a broad international readership. It focuses on practical approaches to teaching physics at the secondary school and early undergraduate levels, publishing peer-reviewed articles that explore novel instructional strategies, experimental techniques, curriculum design, and technological advancements in education.

The journal covers a wide range of topics, including hands-on experiments and demonstrations, the use of technology in physics education, and conceptual understanding along with common student misconceptions. Moreover, the journal highlights additional insights, such as interdisciplinary applications of physics, and historical and philosophical perspectives on physics teaching.

Over the years, *Physics Education* has featured special issues and thematic collections addressing emerging trends in physics education, such as the integration of computational tools, online learning environments, and inquiry-based teaching approaches. The journal has remained a valuable resource for educators worldwide, providing both theoretical insights and practical guidance for classroom implementation. For this study, we had access to *Physics Education*'s entire publication history from its launch in 1966 to 2024.

## 2. 3. LDA theory and prior work

### 2.3.1. Introduction to LDA

Latent Dirichlet Allocation (LDA) is a widely used probabilistic generative model for topic modeling, originally introduced by Blei, Ng, and Jordan in 2003 [21]. It is a technique developed within the field of natural language processing (NLP), a subfield of machine learning that focuses on the recognition and analysis of language and text. LDA is an unsupervised learning method, meaning that it discovers patterns in text without requiring manual labeling or predefined categories [22]. This makes it a powerful tool for uncovering hidden thematic structures in large collections of textual data.

At its core, LDA assumes that each document contains a mix of topics, and each topic is characterised by a distinct set of words with varying probabilities. In other words, when writers use a similar set of words frequently within a text, they are likely discussing the same underlying topic. For example, if a document frequently contains words like "planet", "orbit", "gravity", and "telescope", it is likely discussing astronomy. However, LDA also assumes that documents are not strictly about a single topic; rather, each document is a blend of multiple topics to different degrees. For instance, an article on space exploration may be composed of 70% astronomy-related words and 30% engineering-related words.

LDA functions by representing documents using a "bag of words" approach (BoW), where the order of words is ignored, and only their frequency is considered. The algorithm then tries to group words that frequently co-occur across multiple documents into latent topics. These topics are not predefined but rather inferred from statistical patterns in the data [23]. Once the topics have been identified, LDA outputs them as lists of words weighted by their importance to the topic. Researchers must then interpret these lists to determine the actual meaning of each topic.

One of the most useful aspects of LDA is its ability to calculate topic prevalence across a dataset. By analysing the proportion of each topic in each document, researchers can identify how themes have evolved over time or how they differ between various sources. For example, in an analysis of physics education research, one might observe that discussions about computational methods have increased in recent years, while interest in traditional laboratory-based instruction has declined [24, 25].

While LDA is a powerful tool, it requires researchers to make several decisions, including the number of topics to extract, how much overlap to allow between topics, and how to preprocess the text (such as removing common words like "and" or "listen"). Additionally, because LDA is a probabilistic model, results may vary slightly between runs, requiring careful validation and interpretation [26].

### 2.3.2 Previous applications of LDA to Science and Physics Education Research

Several studies have applied LDA to analyse educational literature, demonstrating its potential for large-scale thematic review, classification and trend analysis. Below, we summarise key works that have utilised this method to study different aspects of science and physics education research.

One of the most comprehensive applications of LDA in science education is the work of Odden, Marin, and Rudolph published in 2021 [27], who used topic modeling to analyse over 100 years of publications in the journal *Science Education*. Their study identified 21 distinct topics in the science education research literature, grouped into three broad categories: science content, teaching-focused, and student-focused topics. By tracking topic prevalence over time, they revealed how national policies, intellectual shifts, and interdisciplinary influences have shaped the field. Their findings suggest that LDA can effectively capture long-term trends and provide insights into the historical development of educational research.

In PER, Odden, Marin, and Caballero in 2020 applied LDA to 18 years of Physics Education Research Conference (PERC) proceedings [25]. They identified ten recurring themes in PER, showing how research interests evolved from early qualitative studies of student understanding to later emphases on problem-solving and, more recently, sociocultural perspectives on teaching and learning. Their work highlights the usefulness of LDA in detecting waves of intellectual focus within a research community.

Building on these efforts, Odden and colleagues in 2024 [28] extended the use of natural language processing for PER-literature analysis by integrating text embeddings with LDA results. This approach allowed them to redefine topic structures and perform a deductive qualitative analysis at scale. By mapping articles into a high-dimensional semantic space, they demonstrated how topic modeling can be augmented with additional machine learning techniques to enhance interpretability and researcher-defined classifications.

## 3. Aims and research questions

While these previous studies have successfully used LDA to examine trends in science and physics education research, they have primarily focused on *research* literature rather than *teaching*-focused journals. Moreover, prior analyses have largely centered on either a single journal (e.g., *Science Education*) or conference proceedings (*PERC*), leaving a gap in understanding how educational discourse has evolved across different journals specifically aimed at physics educators.

In this study, we apply LDA to *The Physics Teacher* (TPT) and *Physics Education* (PE), two long-standing journals dedicated to physics teaching. Unlike previous work, which primarily analysed physics education research literature, our focus is on the thematic evolution of physics teaching practices, curriculum development, key physics contents, and instructional strategies as reflected in these journals. By conducting a comparative analysis of TPT and PE, we aim to uncover how physics teaching priorities and methodologies have changed over time and across different educational communities.

Our research questions are as follows:
1. What topics can be discovered in the corpus of articles composed of all articles published in *The Physics Teacher* and *Physics Education* journal from 1966 through 2019? What characteristic trends about physics teaching can be recognised, how did they evolve, and what is their prominence in this ensemble of articles?
2. To what extent the discovered topics are covered by the two journals separately? Are there specific trends or interests that differentiate the two journals?
3. Considering the reference literature, what do these topics and their trends tell us about the differences and similarities between the two educational systems (European and North American) in regard to the teaching of physics?

# 4. Methodology

To answer these research questions, we performed two analyses:

1. An inductive LDA analysis of the complete dataset, which includes both TPT and PE articles, to identify topics covered by the two journals.
2. A quantitative analysis of the results from the inductive LDA analysis, where we separated the data by TPT and PE articles, to identify differences between the two journals.

It is important to note that before conducting the inductive analysis of the complete dataset, we carried out exploratory LDA analyses on each journal independently. While these individual analyses are not the focus of this work, they informed several decisions in the analysis on the complete dataset. For instance, when determining the number of topics, we considered that each journal independently revealed at least 15 topics, which guided our selection of the minimum number of topics to explore in the combined dataset.

Below, we provide a detailed description of the methodology.

## 4.1. Inductive thematic analyses using LDA

Our general method for topic analysis using LDA consists of three main steps: 1) dataset creation and text preparation, 2) development and testing of the LDA model, 3) topic interpretation and trend visualisation.

### 4.1.1. Dataset creation and text preparation

The creation of the dataset and adjustment of the text is the first crucial step in the analysis process. This phase involves organising and preparing the textual documents, formatting them in a way that can be easily processed by the machine. In fact, the algorithm will use them as input for the development of the LDA model. We note that the TPT and PE articles were retrieved in two different years, so any differences in handling the articles of the two journals will be highlighted.

**PDF-to-text conversion:** we first received access to all articles published in PDF format from both journals. These PDFs were converted into text documents using Adobe Acrobat. The TPT articles were retrieved in mid-2020, and PE articles in late 2024. The extracted text was organised into a structured table, including metadata such as article titles, publication years, authors, and DOIs.

**First dataset cleaning step**: Next, we filtered articles based on relevance for the analysis, removing those primarily focused on journal business (announcement, editorials, and advertisements), assuming that they offer minimal value to our research objectives. Specifically, we removed documents with very short texts (under 400 characters), articles without author information, duplicates, and articles with problematic formatting. Additionally, articles with specific headers, such as "ANNOUNCEMENT", "BOOK REVIEWS", and "NEWS", were excluded.

**Articles exploration:** To confirm proper retrieval of data and information to further refine the dataset, we examined the TPT and PE datasets. First we had a general overview of articles, noting that PE's ones appeared to be better retrieved in terms of titles and formatting than TPT articles. One possible reason for this is that retrieval libraries were improved over the four-year period.

Then, we examined the two datasets to identify specific article series or types to define which should be included in the final dataset. By analyzing the most common bigrams and trigrams in the titles, we identified recurring series such as "NOTES", "Apparatus for teaching physics", and "Research frontier" in TPT, and "PEOPLE", "Reviews", and "Letters" in PE. We also examined the articles' content and noted that some text unrelated to the article itself appeared before or after the main body of the article.

Based on these findings, we proceeded with a second cleaning step.

**Second dataset cleaning step:** Based on the RQs, we included articles relevant to the journals' specific interests (e.g., "Apparatus for Teaching Physics") and excluded those related to journal business or book reviews, maintaining consistency with the first cleaning step. Therefore, we removed articles with specific terms in the titles, such as "Correction", "BOOKS FOR YOUNG PEOPLE", and "Erratum" for TPT, and "News", "PEOPLE", and "book review" for PE.

Moreover, articles missing author information or text, as well as those shorter than 500 words, were excluded, under the assumption that longer articles provide more in-depth treatment of topics.

Finally, since we noticed that each article started with the title (which is not always the case, as in some journals the title is placed at the end or next to the body of text), we filtered out the text preceding each title. This step significantly improved the data, as over 5,000 articles in each journal were corrected at this stage.

This led to the removal of 55 articles from TPT and 252 from PE. Final dataset contained 6445 articles from PE and 7203 from TPT. Therefore, the model was trained on 13648 articles.

**Dataset limitations and impact on the LDA model:** The dataset's limitations stem primarily from challenges with TPT's older articles, particularly those published before digital typesetting. These articles often contained OCR (optical character recognition) errors, which we identified through a random selection process. A common issue was the incorrect merging of sections, as some older PDFs consisted of scanned images rather than selectable text. To estimate the extent of this problem, we grouped articles by year and reviewed a random sample. Our qualitative exploration revealed that articles published before 1978 were most affected by these errors.

These OCR-related issues limit our ability to retrieve text in a more organised form or to remove undesired text that was incorrectly included. Consequently, the quality of the text in these older articles may influence the prevalence of specific word clusters identified by the LDA model, particularly for articles published before 1978.

However, we acknowledge that in a BoW-based model like LDA, the order of words is not a crucial factor. As such, the inclusion of unordered text in the analysis does not undermine the validity of the model's results. Additionally, very short articles were excluded from the dataset, leaving only longer articles where a small amount of anomalous text is unlikely to significantly alter the overall meaning. Given these considerations, we believe that the impact of these limitations is minimal, and the general validity of the analysis remains intact.

**Text preparation for LDA (or LDA pre-processing):** To ensure consistency and reduce the dimensionality of textual data, we applied a series of preprocessing steps. First, unique identifiers and common section headings (e.g., *introduction, methodology, results*) were removed. Next, all newline characters, tabs, digits, and bullet points were eliminated, and non-alphanumeric characters were replaced with spaces. Hyphenated word fragments at line breaks were merged (e.g., *"pre-\nprocessing"* becomes *"preprocessing"*), and the text was converted to lowercase. Finally, minor lexical corrections were applied, such as fixing typos (e.g *tlie* to *the* and *per cent* to *percent*). These steps ensured a clean and standardized text for subsequent analysis.

**Text tokenization and bigrams generation**

To prepare the text for analysis, we applied a series of transformations. First, sentences were tokenized into lowercase word sequences while removing accent marks and punctuation. Next, stopwords (such as "if", "and", "but", etc.) were removed to retain only semantically meaningful terms. Lemmatization was then performed using WordNet, where each token was assigned its most relevant part-of-speech tag (e.g., *"matches"* becomes *"match"*) before being reduced to its base form. Finally, bigrams (frequent co-occurring word pairs) were generated to capture multi-word expressions (for example, converting "high school" to "high_school"), enhancing contextual understanding in the subsequent analysis.

4.1.2. LDA model development and testing

After preparing textual datasets for TPT and PE articles, we used them to develop the LDA model. A series of parameter-tuning steps were implemented to ensure robust and meaningful topic extraction.

**Word frequency filtering:** Before applying the LDA algorithm, we filtered words based on their frequency across the dataset. Words occurring fewer than 15 times were removed to eliminate rare terms that could introduce noise. Additionally, highly frequent words were excluded based on a dynamic threshold, removing terms appearing in anywhere from 45% to 99% of the documents. We systematically analyzed the lists of removed words at different cutoffs to ensure that important domain-specific terms were retained. For instance, setting a stricter threshold at 40% would have resulted in the exclusion of key terms such as *data, measurement, observe, model, force* (from *Physics Education*) and *data, laboratory, motion* (from *The Physics Teacher*).

**Topic extraction using LDA and testing:** Once data preparation and filtering were completed, we combined the two datasets and proceeded with topic modeling using LDA. We used the open-source Gensim library [29] in Python, in combination with the Natural Language Toolkit (NLTK) [30] for data filtering. While the LDA implementation is straightforward in principle, unsupervised topic modeling presents several inherent challenges:

- Random initialization - LDA is a probabilistic model, meaning each run is randomly initialized, which can lead to variations in the extracted topics.
- Number of topics - The number of topics must be specified before running the model. Determining the optimal number of topics is not trivial, as too few topics might obscure meaningful distinctions, while too many might result in overly fragmented themes.

To address these issues, we systematically tested different model configurations to find the optimal number of topics and initial seed. We assumed that the number of topics that could meaningfully describe the dataset, composed of both PE and TPT articles, would be at least 15, because this was the number of topics discovered in each journal when analysed independently. Thus, we tested topic numbers in the range of 15 to 22.

In this phase, we also computed the topic coherence score [31, 32] for each model. Coherence measures how well a model's topics align with actual document content, with typical scores for academic literature ranging from 0.4 to 0.6 [25, 32]. The final selection of 20 topics was made by considering both the elbow method (to compute the best coherence score range) and qualitative inspection (to ensure the interpretability of topics). This approach allowed us to ensure that the extracted themes were distinct and representative of trends across both journals.

To further ensure the robustness of our chosen model, we examined topic stability across different random initializations. Following Odden, Marin, and Caballero (2020) [25], we generated 15 models and used a clustering algorithm to average them, selecting the model closest to this average. This model typically had a higher-than-average coherence score. The final model for the combined PE-TPT analysis achieved a coherence score of 0.577, slightly above the overall average (0.572), which provided additional confidence in its reliability.

### 4.1.3. Topic interpretation and trend visualisation

Once the final LDA model was selected, we conducted a face validity check by manually inspecting the most representative words for each topic to ensure coherence and interpretability. Moreover, we identified and reviewed the top 15 documents most strongly associated with each topic (with topic proportions typically ranging from 60% to 98%). These documents, along with top words, were used to assign meaningful labels to each topic.

Finally, we explored temporal trends to see how topics evolved over time. For this analysis, we calculated their annual prevalence, averaging topic contributions across all documents per year. By "prevalence" we refer to the extent to which each topic is represented in the literature for a given year[1]. To smooth fluctuations, we applied a 3-year rolling average, preserving major trends while reducing noise.

For this analysis, data were considered between the interval from 1966 to 2019. The endpoint, 2019, was selected because data collection for *The Physics Teacher* was conducted in mid 2020, and we included only the years for which the article collection was fully completed. This time frame allows us to observe the trends of these 20 topics from the second part of the 1960s to the pre-pandemic period. 2020 represents the beginning of a new era characterised by significant events such as the massive e-learning and the advent of artificial intelligence tools in schools.

These methodological choices ensured that our LDA model produced interpretable, reliable, and meaningful topic distributions, which were then analysed to construct narratives about trends in the literature.

## 4.2. Quantitative analysis of topic differences between TPT and PE

Based on the previous LDA analysis, each article was represented as a weighted combination of the 20 identified topics. For this analysis, the data were separated into two distinct subsets: one containing only TPT articles and the other containing only PE articles.

---

[1] Prevalence has also been referred to as "attention", denoting the percentage of focus that a particular topic receives within the literature as a whole each year [33].

By plotting the trends of topic prevalence over time from 1966 to 2019, we quantitatively assessed the prominence of each topic within the two journals. This approach facilitated a comparative analysis of the thematic focus of TPT and PE, highlighting potential differences in the interests and trends between the two journals over the study period.

## 5. Results

In this section, firstly we show what emerges by analysing the whole corpus of articles together, in order to have an overall view of the trends in the literature as a whole. Then, in the last section, we present the results of the two separate analyses of articles from *Physics Education* and *The Physics Teacher*, to highlight possible similarities and differences between the journals.

### 5.1. Topics discovered in the inductive LDA analysis (1966-2019)

The analysis of the complete TPT-PE literature in the period 1966-2019 has revealed a diverse set of themes that characterise the research published in *Physics Education* and *The Physics Teacher*. These topics encompass both core physics content and broader themes related to physics pedagogy, instructional strategies, and students' disciplinary learning. For better understanding the results description, in Table 2 we show the complete list of topics discovered, and in Figures 1 and 2, we present the prevalence of the twenty topics in the dataset of articles over time.

The key difference between Figures 1 and 2 lies in how prevalence is represented. Figure 1 displays the *normalized* prevalence of each topic, providing a view of how each topic and topic group evolved over time as a percentage of the complete literature in that year. In contrast, Figure 2 presents the *un-normalized* prevalence, showing the aggregate number of articles published for each topic (summing percent contributions from each article in each year), offering insight into the overall volume of research output.

As listed in Table 2, the first fourteen discovered topics are directly related to physics content, covering key areas such as Newtonian mechanics, waves and sound, electromagnetism, thermodynamics, fluid mechanics, optics, and nuclear physics. These topics reflect the foundational principles and applied aspects of physics, serving as essential components of both high school and university-level instruction. For instance, the **Newtonian mechanics** and **Applied mechanics'** topics explore motion, forces, and energy. Example articles from these topics include "*What is Centrifugal Force?*" (TPT, 1980) [34], which examines how centrifugal force is presented in introductory physics textbooks and its diverse interpretations, and "When equal masses don't balance" (PE, 2004) [35], an article that explains the function of a modified Atwood's machine in an accelerated frame of reference without using calculus, while emphasizing the importance of selecting appropriate systems for free-body diagrams..

Meanwhile, articles featuring the **Circuits and electronics** topic focus on electrical components and their applications, with studies such as "*Using a Multimeter to Study an RC Circuit*" (TPT, 1995) [36] and *Analogue-to-Digital Conversion Made Easy*" (PE, 1992) [37] demonstrating practical techniques for circuit analysis and digital signal processing. Articles within the, **Astronomy** topic delve into celestial mechanics and observational techniques, with studies like "*Solar and Lunar Demonstrators*" (PE, 2009) [38] providing hands-on models for teaching astronomical phenomena, and "*Simulating the Phases of the Moon Shortly After Its Formation*" (TPT, 2014) [39] offering insights into lunar evolution. At the smallest scale, articles on **Fundamental particles and interactions** address the subatomic structure of matter, as explored in *"The Standard Model"* (PE, 1992) [40], which outlines the framework of particle physics, and *"A Temporary Organization of the Subatomic Particles"* (TPT, 1975) [41], which examines early classifications of elementary particles.

These content-focused articles have played a crucial role in the journals by serving as a forum for physics teachers to engage with disciplinary knowledge relevant to their teaching practice. The journals provide space for discussions on high-school-level physics content, often debating finer conceptual points, clarifying common misconceptions, and offering alternative explanations for various phenomena.

Table 2. Topics discovered in the inductive thematic analysis of complete dataset (articles from TPT and PE journals). Content-based topics are blue colored (Topics 1-14), the topic about history and philosophy of physics is green colored (Topic 15), teaching-focused topics are pink colored (Topics 16-19), and the learning focused topic is light brown colored (Topic 20).

| Topic number | Topic name |
|---|---|
| Topic 1 | Fundamental equations and mathematical principles in physics |
| Topic 2 | Newtonian mechanics |
| Topic 3 | Applied mechanics |
| Topic 4 | Fluid mechanics and gas laws |
| Topic 5 | Waves and sound |
| Topic 6 | Colour and light |
| Topic 7 | Optics |
| Topic 8 | Electromagnetism and currents |
| Topic 9 | Electrostatics, and atmospheric physics |
| Topic 10 | Circuits and electronics |
| Topic 11 | Thermodynamics |
| Topic 12 | Astronomy |
| Topic 13 | Nuclear physics and medical physics |
| Topic 14 | Fundamental particles and interactions |
| Topic 15 | History and philosophy of physics |
| Topic 16 | Trends in physics education: course structures and career pathways |
| Topic 17 | Programs, texts, and teaching strategies |
| Topic 18 | Demonstrations and apparatus in physics teaching |
| Topic 19 | Data acquisition and measurement in physics education |
| Topic 20 | Innovative learning strategies in physics education |

Beyond core physics content, **Topic 15, History and philosophy of physics**, explores the lives and contributions of key scientists (e.g. *"Albert Einstein: His Life"*, 1979 - PE [42]), the evolution of scientific thought (*"Assimilation and Development in Arab Scientific Thought"*, 1966 - TPT [43]), and the role of historical texts (*"Early Science Books and Their Women Translators"*, 1998 - TPT [44]). It also examines the intersection of physics with broader cultural and philosophical themes (*"A non-believer looks at physics,* 1987 - PE [45]). Some articles focus on ethical and social issues in science, related to pivotal experiments in the history of physics (*"Where credit is due The Leaning Tower of Pisa Experiment"*, 1972 - TPT [46]). This topic highlights how historical narratives and philosophical reflections shape our understanding of physics today.

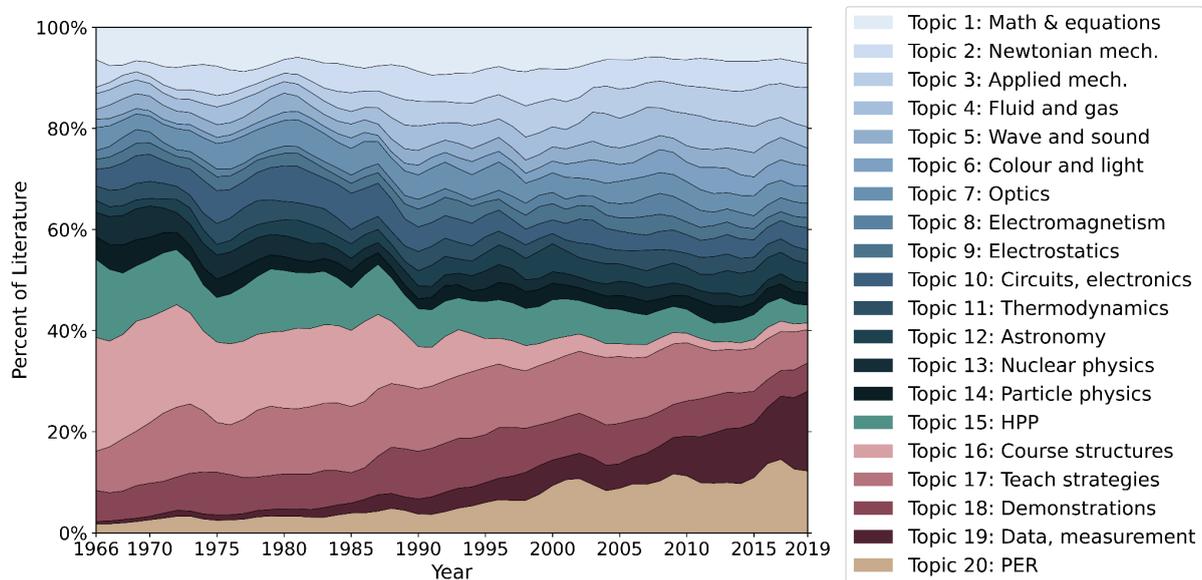

Figure 1: Stacked area plot displaying the normalized prevalence of 20 topics in *Physics Education* and *The Physics Teacher* from 1966 to 2019 (with a three-years rolling window). The plot illustrates the percentage prominence of each topic over time, with content-based topics (blue) at the top, followed by History and philosophy of physics (green), teaching-focused topics (pink), and the PER-focused topic (brown) at the bottom.

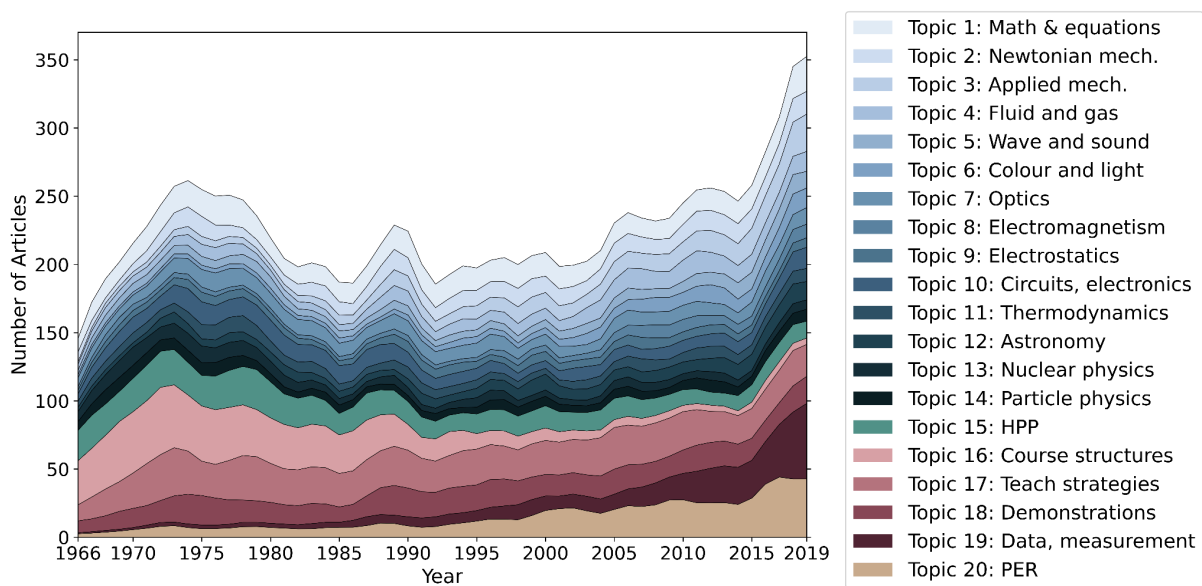

Figure 2: Stacked area plot displaying the un-normalized prevalence of 20 topics in *Physics Education* and *The Physics Teacher* from 1966 to 2019 (with a three-years rolling window). Each topic width represents the number of articles on that topic per year, calculated as the sum of the percent contributions across all articles. Content-based topics (blue) are at the top, followed by the History and philosophy of physics topic (green), teaching-focused topics (pink), and the PER-focused topic (brown) at the bottom.

The next group of topics, 16 to 19, emphasise physics teaching, in the form of pedagogical frameworks and methodologies. The individual topics focus on various aspects of physics teaching, including course structures, career pathways, curriculum design, laboratory instruction, demonstrations, data acquisition, and assessment techniques.

Topic 16, **Trends in physics education: course structures and career pathways** focuses on innovations in the organization and development of physics education courses, examining their relationship with societal and organizational needs, such as careers, occupations, and specializations. The fifteen articles with the highest percent of this topic are all published from the *Physics Education* journal. Example publications include "*Occupation of Successful Candidates in 1969. Graduateship Examination of The Institute of Physics*" (1972) [47], which analyses employment patterns of physics graduates, offering insights into professional trajectories. Alternatively, "*Physics in the polytechnics*" (1980) [48] examines the role of polytechnic institutions in physics education. Similarly, "*The Higher National Diploma in Applied Physics*" (1969) [49] discusses the development of specialized physics programs and their impact on students' technical expertise. These studies highlight the evolution of physics education systems and their alignment with industry and academic demands. As visible in Figure 2, the number of publications under this topic progressively decreased over time, mostly disappearing around the 1990s.

Topic 17, **Programs, texts, and teaching strategies**, most commonly appeared in articles from *The Physics Teacher*, and addresses considerations in school programs, textbook selection, teaching strategies and assessment methods. Example articles include "*Student Selection of the Textbook for an Introductory Physics Course*" (2007) [50] and "*How do you select your physics textbook?*" (1989) [51], which explores how students and instructors navigate the choice of instructional materials, reflecting ongoing discussions about the role of textbooks in physics learning. Similarly, "*Update on the status of the one‑year, non‑calculus physics course*" (1994) [52] reviews the structure and evolution of introductory physics courses, while "'*Retests*'*: a better method of test corrections*" (2011) [53] discusses alternative assessment strategies aimed at improving student understanding through iterative feedback.

Topic 18, **Demonstrations and apparatus in physics teaching,** highlights the development and adaptation of experimental tools to enhance physics instruction. In particular, the topic focuses on hands-on demonstrations designed to make abstract concepts more tangible. Most of the articles that are described by this topic are published in *The Physics Teacher*. For instance, "*Rotational motion demonstrator*" (1988) [54] presents an apparatus to illustrate angular momentum and rotational dynamics in an engaging way, while "*The Iowa wave machines*" (2010) [55] explores an accessible method for visualising wave propagation and interference patterns. Similarly, "*Apparatus for Measuring Young's Modulus*" (2003) [56] provides a straightforward setup for students to experimentally determine the elastic properties of materials. These studies emphasise the crucial role of interactive demonstrations in fostering conceptual understanding and student engagement in physics learning.

Topic 19, **Data acquisition and measurement in physics education**, focuses on the integration of technology and computational tools in experimental physics instruction. This topic appeared in both *The Physics Teacher* and *Physics Education* in articles that highlight innovative approaches to data collection and analysis. For instance, "*Modeling Physics with Easy Java Simulations*" (TPT, 2007) [57] explores the use of computational modeling to visualise and simulate physical phenomena, enhancing students' ability to engage with abstract concepts. Similarly, "*An Arduino Experiment to*

*Study Free Fall at Schools"* (PE, 2018) [58] demonstrates how microcontroller-based experiments can provide accessible, high-precision measurements, fostering a deeper understanding of motion and acceleration. These studies underscore the growing role of digital tools in modern physics education and computational physics, enabling more interactive and data-driven learning experiences.

The final topic, 20, focuses on **Innovative learning strategies in physics education**, representing the intersection between PER and classroom practice. This topic is focused primarily on physics learning, rather than teaching, and examines how research-based insights become crucial to improve physics learning. Results from PER such as active learning and student-centered instruction are advocated to be integrated into teaching to enhance student engagement and learning outcomes. Articles well represented by this topic are *"Using the Force Concept Inventory to Monitor Student Learning and to Plan Teaching"* (PE, 2002) [59], which explores how students' outcomes in concept inventories can inform instructional decisions, as well as, *"Learning from Mistakes: The Effect of Students' Written Self-Diagnoses on Subsequent Problem Solving"* (TPT, 2016) [60] which examines the role of self-reflection in improving problem-solving skills. Another example which critically analyses the impact of textbooks on students' ability to think creatively in physics is *"Physics textbooks: do they promote or inhibit students' creative thinking?"* (PE, 2015) [61].

These discovered topics provide a comprehensive view of the evolving landscape of physics education, reflecting both the interests of the journals and the broader community in key disciplinary content and instructional strategies that shape the teaching and learning of physics.

Furthermore, we found that very few articles are primarily characterized by a single topic. In fact, only 2.8% of the articles are predominantly represented by a topic (with a weight over 75%). The remaining articles typically blend multiple themes. For example, content-based articles may address practical aspects, such as designing effective demonstrations or improving laboratory experiments, thereby reinforcing the connection between theoretical knowledge and hands-on learning.

## 5.2. Major trends in the physics education literature

In this section, we highlight three key periods of increased publication activity, each marking a significant phase in the evolution of physics education.

### 5.2.1. Initial surge in publications (1966-1973)

As illustrated in Figure 2, the annual number of publications increased from approximately 150 to around 250, indicating a consistent upward trend. This can be attributed to the natural growth in the number of publications as both journals became more established and gained recognition in the field. Their increasing reputation likely contributed to a steady rise in submissions over time.

### 5.2.2. Hands-on learning, educational reform, and the rise of PER (1985–1990)

During this period, we observe a notable increase in three specific topics. In particular, topic 18 sees a significant rise in publications about demonstrations and teaching equipment. Based on the historical context, this surge likely reflects a growing emphasis on fostering conceptual understanding. Laboratory demonstrations provided a tangible way to explore physics concepts, an approach further enhanced by the emergence of innovative technological applications during this period. Several articles exemplify this trend: *"A Standing Wave Simulator"* (1989) [62] introduced a tool for visualizing wave behavior, while *"An Interactive Soap Film Apparatus"* (1988) [63] explored novel methods for demonstrating surface tension effects. Additionally, *"The Suitcase Gyroscope. An Angular Momentum Device"* (1987) [64], provided educators with a practical tool to illustrate rotational dynamics, reinforcing the movement toward accessible and impactful teaching demonstrations.

Figure 3: Plot displaying the un-normalized prevalence of specific topics in *Physics Education* and *The Physics Teacher* from 1966 to 2019 (with a three-years rolling window), that are Topic 15 about History and philosophy of physics (light blue), Topic 18 about demonstrations and apparatus (brown), Topic 19 about data and measurement (blue), and Topic 20 on PER (orange). Each topic width represents the number of articles on that topic per year, calculated as the sum of the percent contributions across all articles.

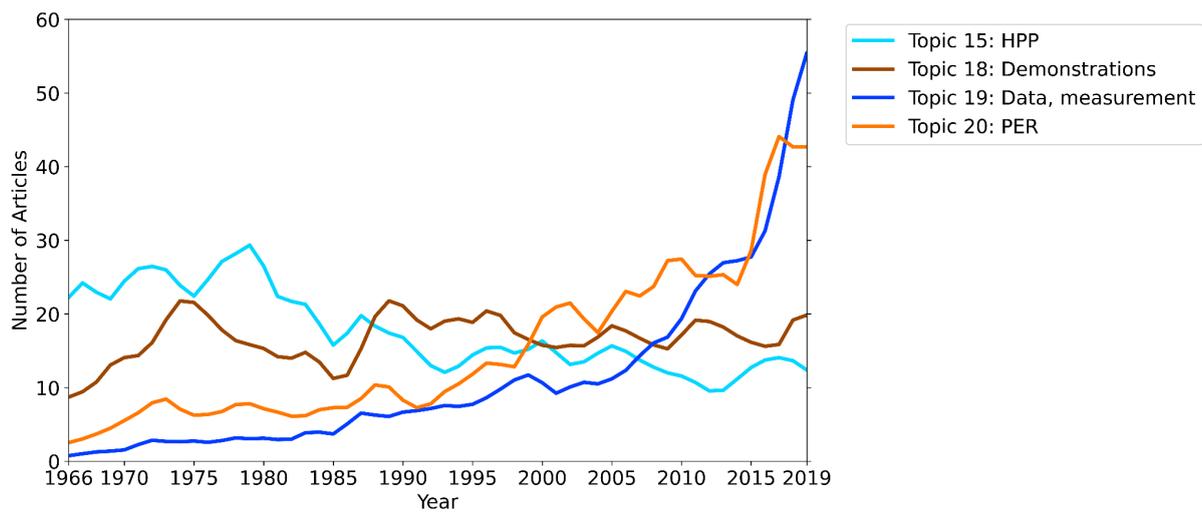

Additionally, during this same time period topic 20 (related to PER) begins to emerge. There were increasing concerns about the effectiveness of physics instruction and the challenges students faced in conceptual understanding. This period coincided with the establishment of PER as area of research, which sought to address these challenges by investigating how students learn physics and developing evidence-based strategies to improve teaching practices For instance, one article from this period is *"An Investigation of Exemplary Physics Teaching"* (Tobin, Deacon, & Fraser, 1989) [65], that highlighted systemic issues in physics education, particularly the dominance of rote memorization over inquiry-based learning. The study echoed broader educational reform movements in the U.S. and U.K., which raised alarms about declining student engagement, shortages of qualified physics teachers, and the lack of real-world relevance in physics curricula. At the same time, researchers and educators were becoming increasingly aware of specific learning difficulties in physics. For example, in *"Student Misconceptions in Mechanics: An International Problem"* (Van Hise, 1988) [66] there is a description of how students persistently held misconceptions about fundamental physics concepts—such as the belief that heavier objects fall faster than lighter ones—even after formal instruction. This article was emblematic of the emerging research paradigm of studying student misconceptions, which became a central focus of PER during this period.

Furthermore, we observe the initial rise of topic 19, which reflects the increasing role of technology in physics education, particularly in data collection and analysis. The late 1980s saw the integration of computers into experimental physics teaching, enabling more precise measurements and real-time data visualization [67]. Articles like *Computer Use for the High School Physics Teacher: Using Available Tool Software* (Irish, 1987) [68] highlight efforts to incorporate computational tools into physics instruction, making data acquisition more efficient. This trend continued into the early 1990s, as seen in *Graphing with Computers in the Physics Lab* (Feulner, 1991) [69], which demonstrated how computer-based graphing tools could enhance students' understanding of experimental results.

5.2.3. Technological integration and PER expansion (2000–2020)

The most significant increase in publications during this period is driven by the growing prominence of topics 19 and 20. The first focuses on computational physics, and the integration of technology in physics teaching, while the second on the influence of PER findings in the teaching activity.

As shown in Figure 3, publications on topic 19 nearly doubled between 2010 and 2020, with a cumulative threefold increase from 2005 to 2020 (Figure 1). Similarly, publications on topic 20, which reflect the growing impact of PER findings on instructional methods and educational policies, began rising in the 1990s and doubled between 2000 and 2020.

Notably, 2005 marked the inception of *Physical Review Special Topics - Physics Education Research* (PRST-PER), later renamed *Physical Review - Physics Education Research* (PRPER). Therefore, this period represents a turning point for the American PER community, as it experienced rapid growth and became an influential field within physics education. The rise of PRPER likely played a direct role in increasing the prevalence of PER-based articles in *The Physics Teacher* (TPT), as many of these publications build on results first published in PRPER.

In these years, multiple factors collectively solidified PER's credibility and impact, fueling its rapid growth [70]. For example, the growing legitimacy within the physics community, as institutions like American Physical Society (APS) formally recognized it as a subfield and tenure-track positions became more available. Additionally, influential assessment tools like the FCI revealed widespread student misconceptions, driving a demand for research-based instructional reforms. Moreover, the expansion of professional development workshops and faculty training programs further disseminated PER findings, while the field's theoretical foundations strengthened through interdisciplinary influences from cognitive science and learning theory.

At the same time, there was a surge in research exploring ways to integrate technology into physics education. This growth was driven by the increasing availability and affordability of computing tools—such as PCs, laptops, smartphones, tablets, sensors, and microcomputers—over the past 10 to 15 years. As these technologies became more accessible, the physics teaching community actively explored their potential to enhance classroom instruction.

Concurrently, the expansion of PER research likely contributed to the growing emphasis on learning strategies, instructional reforms, and the broader integration of research-based teaching practices in physics education.

Table 2. Key trends in physics teaching and publications (1966–2020)

| Period | Focus | Exemplar articles | Historical Context |
| --- | --- | --- | --- |
| 1960s – 1970s | Curriculum Development | - *The Higher National Diploma in Applied Physics* (1969)<br>- *Occupation of Successful Candidates in the 1969 Graduateship Examination of The Institute of Physics* (1972)<br>- *High School Physics Curriculum Development in the Philippines* (1972) | The Cold War and the Space Race drove investments in science education, leading to curriculum reforms such as the PSSC (Physical |

| | | - *The secondary school curriculum and science education* (1973)<br>- *Physics courses at Loughborough* (1978)<br>- *Career oriented pre-technical physics curriculum in San Diego* (1979) | Science Study Committee) and Nuffield Physics. |
|---|---|---|---|
| 1970s – 1990s | Pedagogical Strategies | - *Improving high school physics teacher preparation* (1975) [71]<br>- *Teaching physics: A human endeavor - interview IV* (1977) [72]<br>- *Thinking, reasoning and understanding in introductory physics courses* (1981) [73]<br>- *Optics of the Rear-View Mirror: A Laboratory Experiment* (1986) [74]<br>- *How do you select your physics textbook?* (1989) [75]<br>- *Developing problem-solving thinking in physics education - experience from Papua New Guinea* (1989) [76]<br>- *Removing preconceptions with a "learning cycle"* (1995) [77] | The rise of constructivist learning theories (e.g., Piaget, Vygotsky) emphasized active learning, influencing how physics was taught. |
| 2000s – 2010s | PER and computational physics influence | - *Using the Force Concept Inventory to Monitor Student Learning and to Plan Teaching* (2002)<br>- *Learning from Mistakes: The Effect of Students' Written Self-Diagnoses on Subsequent Problem Solving* (2016)<br>- *Teaching Physics Using PhET Simulations* (2010) [78]<br>- *Understanding resonance graphs using Easy Java Simulations (EJS) and why we use EJS* (2015) [79] | The spread of Physics Education Research (PER) led to data-driven, evidence-based teaching strategies, such as active learning, peer instruction, and computational modeling. |

Table 3. Evolution of laboratory approaches: 1980 vs. 2010

| Period | Main Characteristics | Exemplar articles | Technologies and Methods |
|---|---|---|---|
| 1980s | Traditional approach with analog instruments and mechanical demonstrations. Focus on hands-on experiments with standard equipment. | - *Rotational Motion Demonstrator* (1988)<br>- *The Suitcase Gyroscope. An Angular Momentum Device* (1987) | Use of simple tools such as dynamometers, pendulums, and basic circuits to demonstrate physical principles. Measurements were often taken manually. |
| 2010s | Integration of digital technologies and simulations. Focus on interactive experiments and real-time data analysis. | - *Modeling Physics with Easy Java Simulations* (2007)<br>- *An Arduino Experiment to Study Free Fall at Schools* (2018) | Introduction of digital sensors, Arduino, computer simulations, and data analysis software for more sophisticated and accessible experiments. |

## 5.3. The importance of physics content

We found that a significant number of topics are related to core physics content. Throughout the entire period covered by the two journals, these content-based topics have represented a substantial portion of published articles. The proportion of articles focused on physics content has steadily increased, from approximately 45% of the total publications in 1966 to over 50% by the beginning of 2020 (Figure 1). This upward trend highlights the ongoing prominence of physics content in the literature over time.

While there has been variation in the level of interest in specific topics, with some seeing more attention than others, physics content has remained a central theme. It is important to note, however, that many articles focused on teaching or learning also include substantial physics content, making it challenging to isolate purely content-based articles.

To better understand this distribution, we analyzed the proportion of articles that are predominantly content-focused, finding 118 articles. This means that only 0.95% of all publications can be categorized as purely content-focused, defined as having more than 75% of their topics related to core physics content. Additionally, another small portion of publications (0.61%) have 50% content focus and at least 25% of topics related to teaching, measurement, or PER (either Topics 18, 19, or 20), reflecting a mixed approach that integrates teaching or learning with significant physics content. This highlights that the majority of articles in these journals tend to combine physics content with smaller percentages of many other topics.

## 5.4. Shift from teaching to learning

To better understand how physics education has evolved, we exclude content-based topics (Topics 1–14) and Topic 15, which covers the History and philosophy of physics. This allows us to focus on the development of the remaining topics and uncover broader trends in the field.

When these topics are removed, the normalized distribution of the remaining ones reveals a clear pattern. As shown in Figure 4, what remains are the education-related topics (highlighted in pink), which primarily explore teaching practices, along with Topic 20, which emphasizes the integration of PER findings into classroom instruction. Over time, a noticeable shift emerges: the literature moves away from focusing solely on instructional methods and toward a deeper emphasis on student learning and the impact of research-based teaching strategies.

Specifically, in the 1970s, the predominant focus was on topics such as course structures and career pathways (Topic 16), teaching strategies and programs (Topic 17), and demonstrations and apparatus in physics education (Topic 18). However, by the 1990s and extending into the 2020s, the emphasis shifted towards the integration of data acquisition and measurement in physics education (Topic 19) and innovative learning strategies in physics education (Topic 20).

Figure 4: Stacked area plot showing the normalized prevalence of topics 16 to 20 in *Physics Education* and *The Physics Teacher* from 1966 to 2019 (with a three-years rolling window). This plot illustrates the percentage prominence of each topic over time, highlighting the shift from teaching-focused topics (16-18, pink) to measurement and data (topic 19, purple) and learning-focused topic (topic 20, brown).

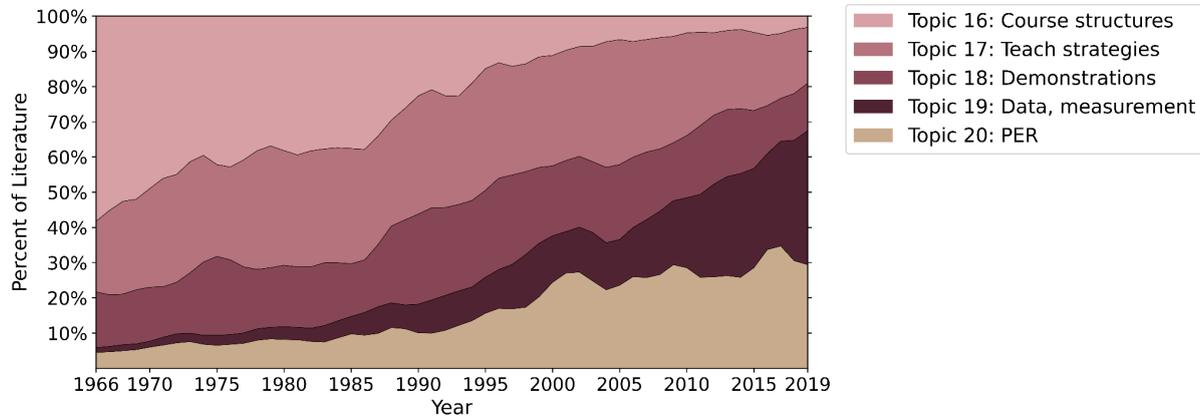

This shift is evident not only in the number of publications specifically on these five topics, but also in the evolving focus of content-based articles. For example, the article *Optics of the Rear-View Mirror* (1986) [74] prioritizes clear content delivery and practical demonstrations for teachers, aiming to improve instruction rather than explore student learning processes. In contrast, a more recent article, *The Image Between the Lenses: Activities with a Telescope and a Microscope* (2003) [80], reflects a later shift towards PER-influenced, student-centered learning by directly addressing students' difficulties in understanding image formation and encouraging inquiry-based exploration. Rather than simply explaining optical principles, this article designs hands-on activities that help students *discover* key concepts, such as the existence of an intermediate real image in telescopes and microscopes.

These articles exemplify the broader transition that gained momentum around 1995, as PER began to influence physics education more strongly. From the mid-1990s onward, both journals increasingly emphasized a research-driven understanding of how students learn physics, sparking new instructional approaches rooted in cognitive science and active learning methodologies.

5.5. Comparison of the two journals

In this section, we present the quantitative analysis of topic prevalence in *The Physics Teacher* and *Physics Education*. By grouping the results of the inductive LDA analysis by journal and examining topics' evolution over time, we compare their thematic focuses.

The stacked area plots in Figures 5 and 6 illustrate the normalized topic trends in each journal from 1966 to 2019, while Figures 7 and 8 present the un-normalized topic prevalence, showing the number of articles dedicated to each topic per year.

Starting with *The Physics Teacher* (Figure 5), a balanced distribution between content-related topics and those centered on teaching practices and strategies is observed. In its early years, the journal gave emphasis on the History and philosophy of physics, a focus that, while still present, has gradually decreased over time. Nevertheless, this early emphasis on the broader context of physics education remains a distinctive feature of TPT.

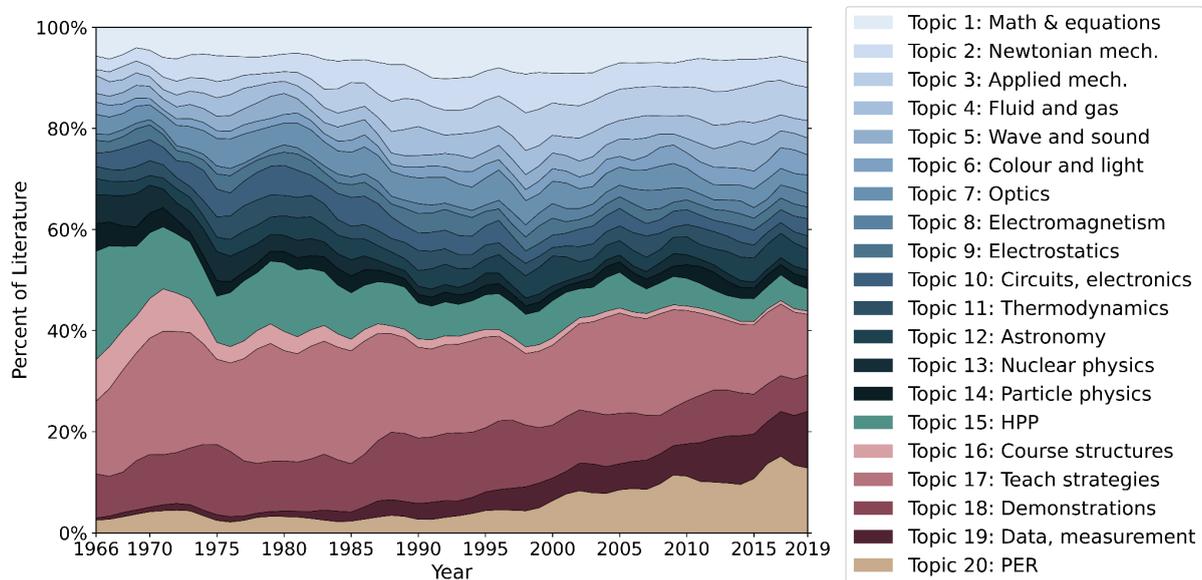

Figure 5: Stacked area plot displaying the normalized prevalence of 20 topics in *The Physics Teacher* from 1966 to 2019 (with a three-years rolling window). The plot illustrates the percentage prominence of each topic over time, with content-based topics (blue) at the top, followed by History and philosophy of physics (green), teaching-focused topics (pink), and the PER-focused topic (brown) at the bottom.

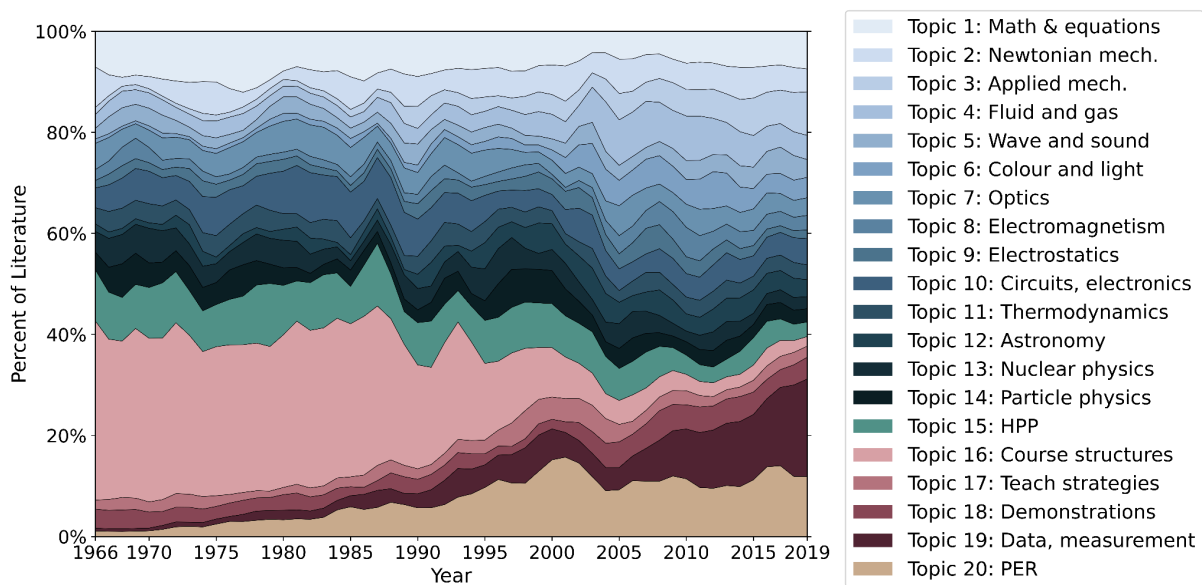

Figure 6: Stacked area plot displaying the normalized prevalence of 20 topics in *Physics Education* from 1966 to 2019 (with a three-years rolling window). The plot illustrates the percentage prominence of each topic over time, with content-based topics (blue) at the top, followed by History and philosophy of physics (green), teaching-focused topics (pink), and the PER-focused topic (brown) at the bottom.

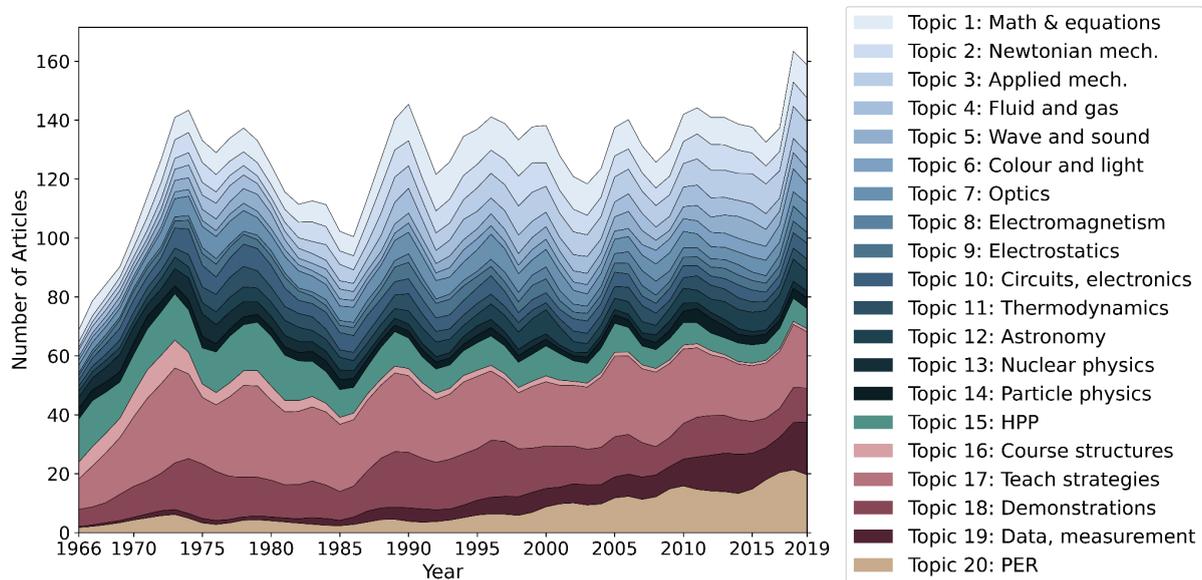

Figure 7: Stacked area plot displaying the un-normalized prevalence of 20 topics in *The Physics Teacher* from 1966 to 2019 (with a three-years rolling window). Each topic width represents the number of articles on that topic per year, calculated as the sum of the percent contributions across all articles. Content-based topics (blue) are at the top, followed by the History and philosophy of physics topic (green), teaching-focused topics (pink), and the PER-focused topic (brown) at the bottom.

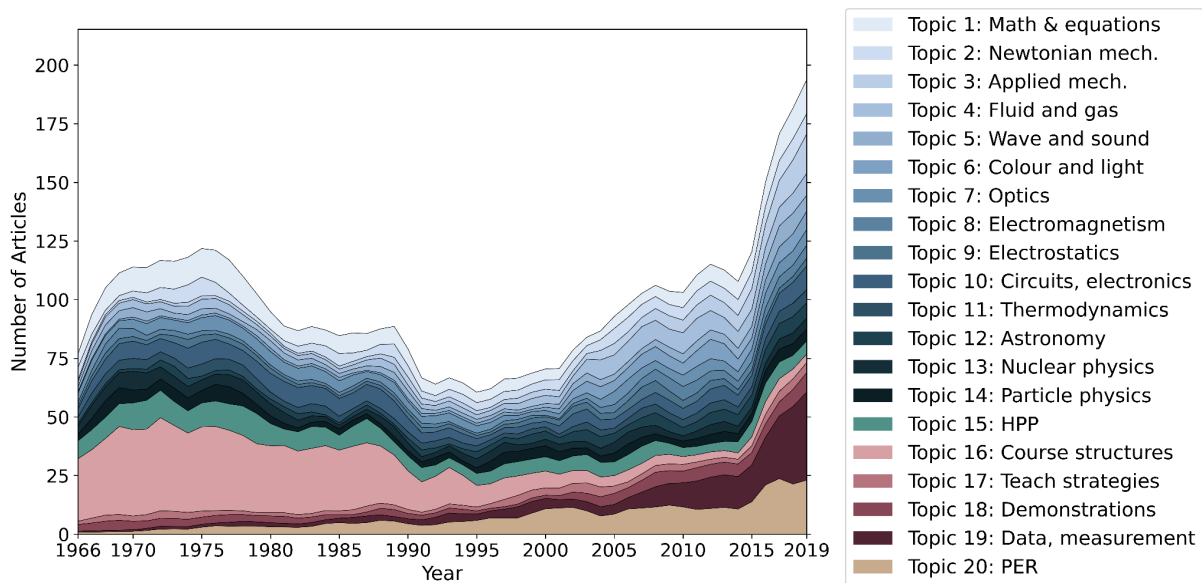

Figure 8: Stacked area plot displaying the un-normalized prevalence of 20 topics in *Physics Education* from 1966 to 2019 (with a three-years rolling window). Each topic width represents the number of articles on that topic per year, calculated as the sum of the percent contributions across all articles. Content-based topics (blue) are at the top, followed by the History and philosophy of physics topic (green), teaching-focused topics (pink), and the PER-focused topic (brown) at the bottom.

Since the 1990s, Topic 19, which relates to data acquisition and measurement, has gained prominence, reflecting technological advancements in physics education. Around 2000, the prominence of Topic 18, which focuses on laboratory apparatus, began to decline, making way for Topic 19. This shift suggests an evolution in the journal's focus from traditional laboratory instruments to a hybrid approach, incorporating modern data acquisition techniques.

Overall, *TPT* has exhibited relative thematic stability, with a few notable shifts, such as the increasing emphasis on PER and educational technology.

In contrast, as for *Physics Education,* Figure 6 reveals a more pronounced evolution in its thematic focus. Initially, from the 1960s to the 1980s, the journal primarily emphasized course structures and curricula. This focus may be attributed to the diversity of educational systems across Europe, where curricula and pathways vary significantly. In contrast, the United States' more standardized educational system may explain why TPT placed less emphasis on curricula and more on general pedagogical strategies. This distinction may warrant further exploration of articles from *PE* published between 1970 and 1985 to understand more about the specific curricula-related content.

A significant shift occurred in the mid-1990s when PE transitioned from curricula-focused discussions to a greater emphasis on content and pedagogy. By the early 2000s, articles on PER increased substantially, continuing to rise throughout the 2010s. This shift paralleled a growing interest in educational technology, aligning with broader trends in physics education. During this period, PE also strengthened its focus on content-based articles, which became increasingly prominent as the pedagogical discussions expanded.

Although PE has maintained some interest in the *History and Philosophy of Physics*, this focus has been notably weaker compared to TPT, with fewer dedicated articles. Instead, PE has increasingly emphasized content-oriented articles, likely responding to the growing demand for pedagogical examples directly tied to physics content. Such articles often combine practical applications with deeper content discussions and teaching methodologies.

A key distinction between the two journals is the timeline of their engagement with PER. PE demonstrated an early interest in PER, beginning in the early 1990s, while TPT shifted towards PER later, in the late 1990s and early 2000s. This earlier focus in PE suggests a proactive engagement with the emerging field of PER, whereas TPT followed broader trends that gained momentum in the 2000s.

Figures 7 and 8 provide additional insight into these trends. In TPT (Figure 7), thematic focus has remained relatively stable, with a modest increase in PER and educational technology. Teaching strategies, textbooks, and demonstrations have persisted as central themes, underscoring TPT's long-standing commitment to practical aspects of physics teaching.

Conversely, PE (Figure 8) exhibits a more dramatic shift. The journal's initial focus on curricula and educational pathways in the 1970s and 1980s gradually gave way to an emphasis on content, pedagogy, and PER by the mid-1990s. The sharp rise in PER-related articles, particularly from 2015 onward, signals a major directional change, aligning with broader transformations in physics education research and practice.

In summary, the analysis highlights key distinctions in the thematic focus of *The Physics Teacher* and *Physics Education*. TPT has maintained a relatively stable balance between content and teaching-related topics, consistently emphasizing pedagogy, history, and philosophy while increasingly incorporating PER and educational technology. In contrast, PE underwent a more pronounced shift, moving away from curricula-focused discussions in the mid-1990s toward greater emphasis on content, pedagogy, and eventually PER and technology in the 2000s and beyond.

# 6. Discussion

In summary, by analyzing the topics covered in *The Physics Teacher* and *Physics Education* over more than five decades, we have identified key themes, their evolution, and the distinct roles these journals have played in shaping physics education discourse. These topics can be grouped into distinct thematic categories, reflecting the evolving priorities and interests in physics education over time.

*Content-based topics*, which have remained consistently present throughout their histories, form a foundational pillar of the literature in both journals. These topics focus on core physics concepts and their instructional representation, serving as a key resource for educators seeking content-specific guidance.

A second major category encompasses *teaching-focused topics*, including discussions on course structures, pedagogical strategies, laboratory activities, and demonstrations. These themes highlight the journals' role in supporting instructors by providing instructional material and methods to enhance the study of physics, and have been present in both journals from their inception until the present day.

Another noteworthy topic emerges from *Physics Education Research* (PER), which has contributed innovative learning strategies informed by empirical studies on how students engage with physics concepts. The increasing trend of this topic underscores a transition toward research-informed teaching methodologies aimed at improving student learning outcomes.

Finally, we have identified a distinct topic related to *historical and philosophical aspects of physics*, which played a prominent role in the early years of both journals. These discussions contextualized physics education within a broader cultural and scientific narrative, emphasizing the development of the discipline and its philosophical implications.

Building on the historical reconstruction presented in the literature review, the Cold War era was marked by significant efforts in both North America and Europe to enhance physics education. This trend is evident in our findings and helps account for the substantial increase in publications from the establishment of both journals until the early 1970s. Throughout the period analyzed, *The Physics Teacher* and *Physics Education* consistently prioritized physics content, addressing a wide range of topics.

Additionally, both journals played a crucial role in providing resources and support for educators, a defining characteristic from the 1970s through the late 20th century. As for European physics education, historically it was shaped by multiple influences and movements. Our analysis reflects this dynamic by highlighting how *Physics Education* dedicated space to discussions on curriculum development, social aspects such as career pathways, and multidisciplinary connections, particularly with engineering. In contrast, *The Physics Teacher* focused more on teaching methodologies and pedagogical tools. This focus may stem from the relatively homogeneous structure of the North American educational system or reflect a deliberate editorial choice by the journal.

Over the past 30 years, physics education has been profoundly shaped by two key developments: the rise of computational tools and the increasing influence of PER on instructional practices. The 1990s marked a turning point, as advances in technology led to the growing use of software tools, sensors, and data analysis techniques in physics instruction. Our analysis reflects this shift, showing the emergence of topics related to measurement and computation within the journals. This transformation highlights a broader movement toward evidence-based teaching, where technological advancements are not only enhancing laboratory experiences but also reshaping pedagogical strategies to prioritize student engagement and conceptual understanding.

Through our thematic analysis, we quantify a significant shift in both journals from a teaching-centered focus, emphasizing course structures, instructional strategies, and laboratory apparatus (as seen in Topics 16, 17, and 18, which address teaching strategies, physics careers, and traditional laboratory setups), to a more student-centered approach. We hypothesize that this transformation has been driven by the growing influence of PER and the integration of educational technologies, both of which have played a crucial role in advancing active learning and evidence-based instructional practices. This shift reflects a broader evolution in physics education, where student engagement, conceptual understanding, and data-driven pedagogical strategies are increasingly prioritized over traditional models of instruction.

This study offers a novel contribution by systematically analyzing the thematic evolution of *The Physics Teacher* and *Physics Education* over an extended period. By examining these journals collectively and individually, we provide a comprehensive overview of the major themes in physics education while also identifying their distinctive areas of focus.

Our findings offer valuable insights into the historical development of the field and build on prior scholarship, such as the work by Otero and Meltzer, by analyzing a much larger body of literature. While their research provides critical perspectives on specific aspects of physics education, our study extends these discussions over a longer timeframe, revealing broad trends that may be less visible in year-to-year analyses. However, our approach remains at a relatively coarse level, offering general insights rather than fine-grained historical narratives.

By placing our findings within the broader historical context of physics education, we highlight how teaching practices have evolved in response to pedagogical research and technological advancements. PER has played a crucial role in shifting the focus from teaching-centered to learning-centered methodologies, and our study reinforces this trajectory by demonstrating how these shifts unfold over longer time scales. Examining these trends over multiple decades enables us to identify significant transformations. For instance, the increasing role of computation and technology in physics education, that may not be as apparent in shorter-term studies

Our study also has several limitations. First, the dataset primarily reflects published literature and does not necessarily capture actual physics teaching practices in classrooms. The journals analyzed serve as important forums for discourse, but the extent to which their content translates into widespread instructional change requires further investigation.

Second, our analysis is constrained by the methodological limitations of LDA, particularly those associated with the BoW representation. These techniques allow for broad thematic identification but lack the ability to capture nuanced contextual meanings within articles. Future studies could incorporate more advanced NLP techniques to refine topic modeling approaches and enhance interpretability.

Additionally, our study does not extend into the most recent developments in physics education, including the impact of the COVID-19 pandemic and the emergence of artificial intelligence (AI) in educational settings. These factors have the potential to reshape physics instruction significantly, and future research should explore how they are reflected in contemporary discourse.

Finally, drawing historical conclusions from this type of analysis presents inherent challenges. Topic modeling allows us to identify overarching trends that should be understood as a model of this body of literature and interpreted, from a combination of the literature base used, the particular mathematical model of text and algorithm employed in the analysis, and the data cleaning, filtering, and modeling decisions made by the authors. A deeper engagement with primary sources, such as editorial statements, and policy documents, would be necessary to construct a more detailed historical narrative [27].

Despite these limitations, our study provides a foundational perspective on the evolution of physics education discourse, offering a structured framework for future research to build upon. By continuing to examine these trends, educators, researchers, and policymakers can better understand the ongoing transformation of physics education and its implications for future practice.

# 7. Conclusions

Over the past six decades, the literature published in *The Physics Teacher* and *Physics Education* has reflected and shaped the evolution of physics education, mirroring broader pedagogical and technological shifts. Although physics content has remained essential to physics education, laboratory activities have evolved, moving from traditional demonstrative apparatus toward more computational and software-based approaches, including interactive simulations and data-driven experimentation.

Our findings reveal a clear transition from a teaching-centered approach, focused on course structures, demonstrations, and instructional strategies, to a learning-centered paradigm. The growing influence of PER has played a key role in this shift, emphasizing student engagement, conceptual understanding, and evidence-based pedagogical practices. These trends underscore the dynamic nature of physics education as it continues to adapt to new research insights, technological advancements, and shifting educational priorities.

Looking ahead, this study opens the door to further research on the evolution of physics education discourse, both through expanded historical analyses and the application of more advanced computational methods. Future work could explore how emerging trends, such as artificial intelligence, online education, and the post-pandemic learning environment, are reshaping the field.

Moreover, as text-mining techniques and generative AI continue to evolve, they offer promising opportunities to uncover deeper patterns in educational literature, generate novel insights, and assist in synthesizing large bodies of work. These tools have the potential to greatly enhance our understanding of how teaching practices and research agendas develop over time. By embracing these methods, physics education researchers can gain new perspectives on the discipline's past, present, and future, enriching our knowledge of how physics is taught and learned worldwide.

# Acknowledgments

This work began during Martina Caramaschi's PhD period abroad at the University of Oslo's Center for Computing in Science Education, where she worked under the supervision of Tor Ole B. Odden. The authors gratefully acknowledge Olivia Levrini for supervising the PhD research and for providing valuable insights and feedback on this work. Financial support for the research stay was provided by the PhD program in Data Science and Computation of the University of Bologna and the Marco Polo project of the same university.

## Data availability statement

The code and some of the data that support the findings of this article are openly available. Specifically, the article metadata, topics weights, and the article text representation as a bag-of-word (including bigrams) are public on a [GitHub repository](#). The original articles' texts cannot be made publicly available due to copyright.

## References

1. Virgil E. Bottom, Wallace A. Hilton, John I. Lodge, Vernon L. Long, Edwin J. Schillinger, Harold E. Way; Report of the Denver Conference on Curricula for Undergraduate Majors in Physics. *Am. J. Phys*. 1 March 1962; 30 (3): 153–162. https://doi.org/10.1119/1.1941961
2. Recommendations of the Second Ann Arbor Conference on Undergraduate Curricula for Physics Majors. *Am. J. Phys*. 1 May 1963; 31 (5): 328–335. https://doi.org/10.1119/1.1969506
3. E. Leonard Jossem; Undergraduate Curricula in Physics: A Report on the Princeton Conference on Curriculum S. *Am. J. Phys*. 1 June 1964; 32 (6): 491–497. https://doi.org/10.1119/1.1970743
4. David E. Meltzer, Valerie K. Otero; A brief history of physics education in the United States. *Am. J. Phys*. 1 May 2015; 83 (5): 447–458. https://doi.org/10.1119/1.4902397
5. Science for All at CERN" Hof, B. Physics Today 77 (5), 44–51 (2024) https://doi.org/10.1063/pt.pxip.pgns
6. Lewis, J. *Chapter 9. Physics Education*" in "*125 years: the Physical Society and the Institute of Physics*" ed. by John L. Lewis, Institute of Physics Publishing, 1999 XIII, 243 p. : ill. ; ISBN 0-7503-0609-2
7. Richard, Gunstone. PHYSICS EDUCATION PAST, PRESENT AND FUTURE: AN INTERPRETATION THROUGH CULTURAL CONTEXTS. Teaching and Learning of Physics in Cultural Contexts, pp. 25-45 (2004). https://doi.org/10.1142/9789812702890_0003
8. Arons, A. B. (1976). Cultivating the capacity for formal reasoning: Objectives and procedures in an introductory physical science course. Am. J. Phys. 44 (9): 834–838. https://doi.org/10.1119/1.10299
9. McDermott, L. C., Rosenquist, M. L., & van Zee, E. H. (1987). Student difficulties in connecting graphs and physics: Examples from kinematics. Am. J. Phys. 55 (6): 503–513. https://doi.org/10.1119/1.15104
10. Halloun, I. A., & Hestenes, D. Common sense concepts about motion (1985). American Journal of Physics, 53, 1043–1055.
11. Hestenes, D., Wells, M., Swackhamer, G. Force concept inventory (1992) The Physics Teacher 30: 141–166.
12. Weber, J, &. Wilhelm, T. (2020). The benefit of computational modelling in physics teaching: a historical overview. Eur. J. Phys. 41 034003
13. Hake, R. (1998). Interactive-Engagement Versus Traditional Methods: A Six-Thousand-Student Survey of Mechanics Test Data for Introductory Physics Courses. American Journal of Physics - AMER J PHYS. 66. 10.1119/1.18809.
14. McDermott, L. C., & Redish, E. F. (1999). Resource Letter: PER-1: Physics Education Research. *Am. J. Phys*. 67 (9): 755–767. https://doi.org/10.1119/1.19122
15. Knecht, W. New trends in physics teaching, v.1, 1965-1966. Paris: UNESCO, 1968
16. Bevilacqua, F., Giannetto, E. The history of physics and European physics education. *Science and Education* 5, 235–246 (1996). https://doi.org/10.1007/BF00414314
17. May. J. M. (2023) Historical analysis of innovation and research in physics instructional laboratories: Recurring themes and future directions. Phys. Rev. Phys. Educ. Res. 19, 020168. DOI: https://doi.org/10.1103/PhysRevPhysEducRes.19.020168
18. MacDonald, W. M., Redish, E. F., and Wilson, J. M. (1988). The M.U.P.P.E.T. Manifesto, Computers in Physics 2, 23

19. Wilson, J. M., & Redish, E. F. (1989). USING COMPUTERS IN TEACHING PHYSICS, Physics Today 42, 34
20. American Association of Physics Teachers; Goals of the Introductory Physics Laboratory. *Am. J. Phys*. 1 June 1998; 66 (6): 483–485. https://doi.org/10.1119/1.19042
21. Blei, D. M., Ng, A. Y., & Jordan, M. I. Latent Dirichlet allocation. *J. Mach. Learn. Res*. **3** (2003) 993–1022. http://www.jmlr.org/papers/volume3/blei03a/blei03a.pdf
22. Hoffman, M. D., Blei, D. M., & Bach, F. (2010). Online learning for latent Dirichlet allocation. In J. D. A. Lafferty, C. K. I. Williams, J. Shawe‑Taylor, & R. S. Zemel (Eds.), Advances in neural information processing systems 23 (pp. 856–864). https://proceedings.neurips.cc/paper/2010/hash/71f6278d140af599e06ad9bf1ba03cb0-Abstract.html
23. Zhang, Y., Jin, R., & Zhou, Z. H. (2010). Understanding bag‑of‑words model: A statistical framework. International Journal of Machine Learning and Cybernetics, 1(1), 43–52. https://doi.org/10.1007/s13042-010-0001-0
24. Griffiths, T. L., & Steyvers, M. (2004). Finding scientific topics. Proceedings of the National Academy of Sciences of the United States of America, 101(suppl 1), 5228–5235. https://doi.org/10.1073/pnas.0307752101
25. Odden, T. O. B., Marin, A. & Caballero, M. D. Thematic analysis of 18 years of physics education research conference proceedings using natural language processing. *Phys. Rev. Phys. Educ. Res*. **16** (2020) 010142, 1-25 doi: 10.1103/PhysRevPhysEducRes.16.010142
26. Denny, M. J., & Spirling, A. (2018). Text preprocessing for unsupervised learning: Why it matters, when it misleads, and what to do about it. Political Analysis, 26(2), 168–189. https://doi.org/10.1017/pan.2017.44
27. Odden, T. O. B., Marin, A. & Rudolph J. L. How has Science Education changed over the last 100 years? An analysis using natural language processing. *Sci. Ed*. **105** (2021) 653–680. doi: 10.1002/sce.21623
28. Odden, T. O. B., Tyseng, H., Mjaaland, J. T., Kreutzer, M. F. & Malthe-Sørenssen, A. (2024) Using text embeddings for deductive qualitative research at scale in physics education. *Phys. Rev. Phys. Educ. Res*. **20**, 02015. doi: 10.1103/PhysRevPhysEducRes.20.020151
29. Rehurek, R., & Sojka, P. (2011). Gensim–python framework for vector space modelling. NLP Centre, Faculty of Informatics, Masaryk University.
30. Bird, S., Klein, E., & Loper, E. (2009). Natural language processing with Python: Analyzing text with the natural language toolkit. O'Reilly Media, Inc
31. Röder, M., Both, A., & Hinneburg, A. (2015). Exploring the space of topic coherence measures. WSDM 2015—Proceedings of the 8th ACM International Conference on Web Search and Data Mining (pp. 399–408). https://doi.org/10.1145/2684822. 2685324
32. Syed, S., & Spruit, M. (2018). Full‑text or abstract? Examining topic coherence scores using latent Dirichlet allocation. In 2017 IEEE International conference on data science and advanced analytics (DSAA) (pp. 165–174). https://doi.org/10. 1109/DSAA.2017.61
33. Jacobi, C., Van Atteveldt, W., & Welbers, K. (2016). Quantitative analysis of large amounts of journalistic texts using topic modelling. Digital Journalism, 4(1), 89–106. https://doi.org/10.1080/21670811.2015.1093271
34. Bauman, R. P. (1980). What is centrifugal force? *The Physics Teacher*, 18(7), 527–529.
35. Newburgh, R., Peidle, J. and Rueckner, W. (2004). When equal masses don't balance. Phys. Educ. 39, 3, 289. doi: 10.1088/0031-9120/39/3/006
36. John G. King and A. P. French. (1995). Using a multimeter to study an RC circuit. *The Physics Teacher*, 33, 3, 188–189. 10.1119/1.2344189
37. Jenkins, T. E., and Goodes, S. R. (1992) Analogue-to-digital conversion made easy *Phys. Educ*. 27 116
38. Ros, R. M. (2009). Solar and Lunar Demonstrators Physics Education, 44, 4, 345-355
39. Noordeh, E., Hall, P., Cuk, M. (2014). Simulating the phases of the Moon shortly after its formation. *The Physics Teacher*, 52, 4, 239-240
40. Ryder, L. H. (1992). The Standard Model. *Physics Education*, 27(1), 29–32.

41. Swartz, C. E. (1975) A temporary organization of the subatomic particles *Phys. Teach.* 13, 542–544 https://doi.org/10.1119/1.2339260
42. Rosser, W. G. V. (1979) *Albert Einstein: His Life Phys. Educ.* 14 220
43. Revere, R. B. (1966) Assimilation and Development in Arab Scientific Thought *Phys. Teach.* 4, 350–353 https://doi.org/10.1119/1.2351046
44. Badilescu, S. (1998) *Early Science Books and Their Women Translators. Phys. Teach.* 36, 516–518 https://doi.org/10.1119/1.880123
45. Hermann, B. (1987). *A non-believer looks at physics. Phys. Educ.* 22 280 DOI 10.1088/0031-9120/22/5/316
46. Raman, V. V. (1972) *Where credit is due* The Leaning Tower of Pisa Experiment. *Phys. Teach.* 10, 196–198 https://doi.org/10.1119/1.2352161
47. C Greaves, C. (1972) Occupation of successful candidates in the 1969 graduateship examination of The Institute of Physics. *Phys. Educ.* 7 115 DOI 10.1088/0031-9120/7/2/314
48. Pointon, A. J. (1980) Physics in the polytechnics *Phys. Educ.* 15, 1, 37. DOI 10.1088/0031-9120/15/1/310
49. Hughes, D. O. and Teale, R. (1969) The Higher National Diploma in Applied Physics. *Phys. Educ.* 4, 5, 30. DOI 10.1088/0031-9120/4/5/311
50. Dake, L.S. (2007). Student Selection of the Textbook for an Introductory Physics Course" *The Physics Teacher*, 45, 7, 416–419. DOI: 10.1119/1.2783148
51. Schultz, F. H. C. (1989) How do you select your physics textbook?. *The Physics Teacher*, 27, 4, 278. DOI: 10.1119/1.2342758
52. George, S. (1994) Update on the status of the one-year, non-calculus physics course. *The Physics Teacher*, 32, 5, 344–346. DOI: 10.1119/1.2343097
53. Hake, E. J. (2011). 'Retests': A better method of test corrections. *The Physics Teacher*, 49, 3, 168–171. DOI: 10.1119/1.3555393
54. Pizzo, J. (1988). Rotational motion demonstrator. *Phys. Teach.* 1 March 1988; 26 (3): 187–188. https://doi.org/10.1119/1.2342473
55. Daffron, J. D., Greenslade, T. B. and Stille, D. (2010). The Iowa wave machines. *Phys. Teach.* 48, 3, 200–201. https://doi.org/10.1119/1.3317460
56. Niculescu, A. and Shumaker, R. (2003). Apparatus for Measuring Young's Modulus. *Phys. Teach.* 41, 6, 364–367. https://doi.org/10.1119/1.1607808
57. Christian, W. and Esquembre, F. (2007) Modeling Physics with Easy Java Simulations. *Phys. Teach.* 45, 8, 475–480. https://doi.org/10.1119/1.2798358
58. Moya, A. A. (2018). An Arduino experiment to study free fall at schools. Physics Education. 53, 5. 10.1088/1361-6552/aad4c6
59. Savinainen, A. and Scott, P. (2002). Using the Force Concept Inventory to monitor student learning and to plan teaching. Physics Education, 37, 1, 53. DOI 10.1088/0031-9120/37/1/307
60. Mason, A., Yerushalmi, E., Cohen, E. and Singh, C. (2016). Learning from Mistakes: The Effect of Students' Written Self-Diagnoses on Subsequent Problem Solving. *Phys. Teach.* 54, 2, 87–90. https://doi.org/10.1119/1.4940171
61. Klieger, A. and Sherman, G. (2015). Physics textbooks: do they promote or inhibit students' creative thinking. Physics Education, 50, 3, 305, DOI 10.1088/0031-9120/50/3/305
62. March, R. H. (1989). A standing wave simulator. *Phys. Teach.* 27 (5): 400–401. https://doi.org/10.1119/1.2342809
63. Pizzo, J. (1988). An interactive soap film apparatus. *Phys. Teach.* 26 (4): 238–239. https://doi.org/10.1119/1.2342503
64. Connolly, W., & Koser, J. F. (1987). The suitcase gyroscope—An angular momentum device. *Phys. Teach.* 25 (4): 231–232. https://doi.org/10.1119/1.2342230
65. Tobin, K.,Deacon, J., & Fraser, B (1989). An investigation of exemplary physics teaching. *Phys. Teach.* 27 (3): 144–150. https://doi.org/10.1119/1.2342700
66. Van Hise, Y. A. (1988). Student misconceptions in mechanics: An international problem?. *Phys. Teach.* 26 (8): 498–502. https://doi.org/10.1119/1.2342598
67. Wilson, J. M., & Redish, E. F. (1989). USING COMPUTERS IN TEACHING PHYSICS, Physics Today 42, 34

68. Irish, P. (1987). Computer use for the high school physics teacher: Using available tool software. Phys. Teach. 25 (5): 272–275. https://doi.org/10.1119/1.2342251
69. John Feulner, J. (1991). Graphing with computers in the physics lab. Phys. Teach. 29 (2): 126–127. https://doi.org/10.1119/1.2343242
70. Cummings, K. (2011). *A Developmental History of Physics Education Research*. Retrieved March 31, 2025, from https://sites.nationalacademies.org/
71. McDermott, L. C. (1975). Improving high school physics teacher preparation. Phys. Teach. 13 (9): 523–529. https://doi.org/10.1119/1.2339256
72. Huegel, D. (1977). Teaching physics: A human endeavor Interview IV. Phys. Teach. 15 (9): 538–541. https://doi.org/10.1119/1.2339763
73. Arons, A. (1981). Thinking, reasoning and understanding in introductory physics courses. Phys. Teach. 19 (3): 166–172. https://doi.org/10.1119/1.2340737
74. Jones, E. R., & Edge, R. D. (1986). Optics of the rear‑view mirror: A laboratory experiment. Phys. Teach. 24 (4): 221–223. https://doi.org/10.1119/1.2341989
75. Schultz, F. H. C. (1989) How do you select your physics textbook?. The Physics Teacher, 27, 4, 278. DOI: 10.1119/1.2342758
76. Marshall, S. & Gilmour, M. (1989). Developing problem-solving thinking in physics education - experience from Papua New Guinea. Physics Education, 24, 1, 22. DOI 10.1088/0031-9120/24/1/308
77. Gang, S. (1995). Removing preconceptions with a ''learning cycle''. Phys. Teach. 33 (6): 346–354. https://doi.org/10.1119/1.2344236
78. Wieman, C. E., Adams, W. K., Loeblein, P., & Perkins, K. K. (2010). Teaching Physics Using PhET Simulations. Phys. Teach. 48 (4): 225–227. https://doi.org/10.1119/1.3361987
79. Wee, L. K., Lee, T., Chew, C., Wong, D., & Tan, S. (2015). Understanding resonance graphs using Easy Java Simulations (EJS) and why we use EJS. Physics Education, 50(2) DOI: 10.1088/0031-9120/50/2/189
80. Cox, A. J., & DeWeerd, A. J. The Image Between the Lenses: Activities with a Telescope and a Microscope. *Phys. Teach.* 1 March 2003; 41 (3): 176–177. https://doi.org/10.1119/1.1557509

# Appendix

List of 20 topics discovered in the combined analysis:

| Article title | year | journal |
|---|---|---|
| **Topic 1 - Fundamental equations and mathematical principles in physics:** 2,9%"equation" + 1,0%"constant" + 0.8%"unit" + 0,8%"mass" + 0,7%"term" + 0,6%"equal" | | |
| Plane angle and pi as physical quantities | 1973 | PE |
| Graphical analysis of electric fields of dipoles and bipoles | 2000 | TPT |
| Back to the physics: Integration by Gauss | 1998 | TPT |
| The twin paradox | 1985 | PE |
| Electric field of a two-charge dipole: A graphical approach extended | 2000 | TPT |
| Gravitational Force Due to a Sphere: A Noncalculus Calculation | 2003 | TPT |
| On the teaching of the electric dipole | 2001 | TPT |
| The estimation of errors... | 1966 | PE |
| An alternative method for solving the elementary equation for damped oscillation (teaching) | 1979 | PE |
| A closer look at Fermat's principle | 1986 | PE |
| Perverse symbolism | 1989 | PE |
| The Physics of $\lim_{\theta \to 0}(\sin\theta)/\theta = 1$ | 2019 | TPT |

| | | | |
|---|---|---|---|
| | Moment of Inertia by Differentiation | 2015 | TPT |
| | The vertical spring‑mass system and its "equivalent" | 1976 | TPT |
| | "Bad" vibes: The damped oscillator | 1992 | TPT |
| **Topic 2 - Newtonian Mechanics:** 5,7%"force" + 2,4%"motion" + 1,9%"acceleration" + 1,9%"mass" + 1,8%"object" + 1,7%"body" | | | |
| | Acceleration = | 1966 | PE |
| | What is centrifugal force? | 1980 | TPT |
| | Drawing forces | 1998 | TPT |
| | A very persistent mistake | 2011 | PE |
| | Centrifugal force | 1975 | PE |
| | "Apparent Weight": A Concept that Is Confusing and Unnecessary | 2010 | TPT |
| | NOTES: Right way up | 1973 | TPT |
| | Centrifugal force | 1973 | PE |
| | Do centrifugal forces exist? | 1971 | PE |
| | Newton's laws: a very persistent consistency | 2012 | PE |
| | Physics that textbook writers usually get wrong: III. Forces and vectors | 1992 | TPT |
| | The physics we want | 1985 | PE |
| | When equal masses don't balance | 2004 | PE |
| | Muscle forces | 1992 | PE |
| | Physics in a bouncing car | 1984 | TPT |
| **Topic 3 - Applied Mechanics:** 3,7%"ball" + 2,2%"speed" + 1,5%"velocity" + 1,1%"collision" + 1,0%"fall" + 0,9%"motion" | | | |
| | Spinning eggs and ballerinas | 2013 | PE |
| | Enhancing the Bounce of a Ball | 2010 | TPT |
| | Bounce of a perfectly elastic ball | 2019 | PE |
| | Rolling Motion of a Ball Spinning About a Near-Vertical Axis | 2012 | TPT |
| | Why some balls spin faster than others when they bounce | 2019 | PE |
| | Newton's cradle in billiards and croquet | 2016 | PE |
| | Behaviour of a bouncing ball | 2015 | PE |
| | Ball collision experiments | 2015 | PE |
| | Elastic and Inelastic Collisions of a Ball with a Wood Block | 2017 | TPT |
| | Motion of a Ball on a Moving Surface | 2016 | TPT |
| | Surprising Behavior of Spinning Tops and Eggs on an Inclined Plane | 2016 | TPT |
| | A spinning top for physics experiments | 2019 | PE |
| | Precession of a Spinning Ball Rolling Down an Inclined Plane | 2015 | TPT |
| | Why low bounce balls exhibit high rolling resistance | 2015 | PE |
| | Static friction on a ball rolling down an incline | 2018 | PE |
| **Topic 4 - Fluid mechanics and gas laws**: 4,6%"water" + 2,5%"pressure" + 2,2%"air" + 1,8%"liquid" + 1,4%"volume" + 1,1%"tube" | | | |
| | The Cartesian diver, surface tension and the Cheerios effect | 2014 | PE |
| | Mariotte Bottle with Side Openings | 2006 | TPT |
| | Archimedes' principle in action | 2007 | PE |
| | The secret siphon | 2011 | PE |
| | Filling or draining a water bottle with two holes | 2016 | PE |
| | Simple buoyancy demonstrations using saltwater | 1989 | TPT |
| | Easy test calculates air's weight | 2009 | PE |
| | Liquid nitrogen turns down the temperature on gas-law demos | 2004 | PE |
| | Having fun with liquid nitrogen | 2005 | PE |
| | Fizziology | 2004 | PE |
| | Water barometer | 1993 | TPT |
| | Siphonic concepts examined: a carbon dioxide gas siphon and siphons in vacuum | 2011 | PE |
| | Rubber balloons, buoyancy and the weight of air: a look inside | 2009 | PE |
| | Physics trick gets students interested | 2008 | PE |
| | Measuring liquid density using Archimedes' principle | 2006 | PE |
| **Topic 5 - Waves and sound**: 4,7%"frequency" + 3,6%"wave" + 3,1%"sound" + 1,4%"signal" + 1,0%"amplitude" + 0,8%"pulse" | | | |
| | Using musical intervals to demonstrate superposition of waves and Fourier analysis | 2013 | PE |
| | Another Look at Combination Tones | 2019 | TPT |
| | Fourier Analysis of Musical Intervals | 2008 | TPT |
| | Twenty-four tuba harmonics using a single pipe length | 2017 | PE |
| | Amplitude, frequency, and timbre with the French horn | 2018 | PE |
| | Experimenting with musical intervals | 2003 | PE |
| | Acoustics of percussion instruments-Part I | 1976 | TPT |
| | Teaching Resonance and Harmonics with Guitar and Piano1 | 2014 | TPT |
| | Signal Frequency Spectra with Audacity® | 2015 | TPT |
| | Physics an psychophysics of high‑fidelity sound | 1979 | TPT |
| | Inharmonic Spectra with a Rock Guitar Effects Pedal | 2018 | TPT |
| | Complexity of pitch and timbre concepts | 1998 | PE |
| | Mouthpiece and Bell Effects on Trombone Resonance | 2014 | TPT |
| | Phase shifting and the beating of complex waves | 2011 | PE |
| | Students dance longitudinal standing waves | 2017 | PE |
| **Topic 6 - Colour and light**: 4,9%"light" + 1,8%"image" + 1,5%"colour" + 1,5%"color" + 1,2%"spectrum" + 1,1%"camera" | | | |
| | Yellow: The Magic Color | 2008 | TPT |
| | Demonstrating Fluorescence with Neon Paper and Plastic | 2015 | TPT |
| | A photoshoot for food and drink: camera 'sees' more than you think | 2004 | PE |
| | Yellow | 1994 | TPT |
| | Observing colours and spectra produced by a digital projector | 2007 | PE |
| | Color reproduction with a smartphone | 2013 | TPT |
| | Optical experiments using mini-torches with red, green and blue light emitting\ndiodes | 2007 | PE |
| | Fluorescence Spectra of Highlighter Inks | 2018 | TPT |
| | Color mixing with four prisms redux | 2018 | TPT |
| | Low-cost experiments on infrared phenomena | 2013 | PE |
| | Tricks with invisible light | 2003 | PE |
| | Ultraviolet photography and insect vision | 2019 | TPT |
| | Easy observation of infrared spectral lines | 2012 | PE |
| | Confusing color concepts clarified | 1999 | TPT |
| | "Can You See Me Now?" Using a favorite haunted physics laboratory project to teach color theory | 2011 | TPT |
| **Topic 7 - Optics**: 2,6%"light" + 2,6%"image" + 2,1%"lens" + 1,5%"mirror" + 1,3%"object" + 1,3%"ray" | | | |
| | Virtual Mirrors | 2010 | TPT |
| | Optics of the rear‑view mirror: A laboratory experiment | 1986 | TPT |
| | Multiple images in plane mirrors | 1982 | TPT |
| | Comment on 'From the pinhole camera to the shape of a lens: the camera-obscura reloaded' | 2016 | PE |
| | Camera optics | 1982 | TPT |
| | Polarisation experiments with a Fresnel biprism | 1986 | PE |
| | A simple interferometer | 1997 | TPT |
| | Curved mirrors | 1995 | PE |
| | Demonstration of lens aberrations using a school laser | 1976 | PE |
| | Optical transforms of the alphabet | 1977 | TPT |
| | Lenses, pinholes, screens, and the eye | 1991 | TPT |
| | Refractive index measurement with a ruler | 1984 | TPT |
| | Reflections from a Fresnel lens | 2005 | PE |
| | Reflections on Handedness | 2004 | TPT |
| | Covering lenses and covering images | 1998 | TPT |
| **Topic 8 - Electromagnetism and currents:** 4,2%"magnet" + 3,0%"magnetic_field" + 2,9%"field" + 2,9%"coil" + 2,4%"current" + 2,3%"magnetic" | | | |
| | Universal 'imaginary closed circuit method' and formula for determination of direction of\ninduced emf/current | 2011 | PE |
| | Demonstrate Lenz's law with an aluminium ring | 2008 | PE |
| | Demonstrating the Meissner effect and persistent current | 2000 | TPT |
| | A simple demonstration of motional electromotive force | 2012 | PE |
| | A New Version of an Old Demonstration Experiment Using the Elihu Thomson Jumping Ring Apparatus | 2016 | TPT |
| | Thomson's Jumping Ring Over a Long Coil | 2018 | TPT |
| | Interacting Compasses | 2009 | TPT |
| | A simple way to teach magnetic braking | 2007 | PE |
| | Superconducting cylinders aid in an understanding of current induction | 2004 | PE |
| | Using LEDs to demonstrate \ninduced current | 1992 | PE |
| | A Pictorial Approach to Lenz's Law | 2018 | TPT |
| | Abstract concepts come to life | 2008 | PE |
| | Motional Mechanisms of Homopolar Motors & Rollers | 2009 | TPT |
| | Aluminium rod rolls in non-uniform magnetic field | 2012 | PE |
| | Measuring Earth's Local Magnetic Field Using a Helmholtz Coil | 2014 | TPT |
| **Topic 9 - Electrostatics, and atmospheric Physics**: 2,2%"charge" + 1,6%"surface" +0,9%"plate" + 0,7%"wind" + 0,7%"air" + 0,7%"area" | | | |
| | Suppression of lightning by chaff seeding | 1975 | TPT |
| | The THUNDERSTORM | 1968 | PE |
| | Thermal patterns in the snow. Part II | 1976 | TPT |
| | Exploration glaciology: radar and Antarctic ice | 2007 | PE |
| | Investigation of the airflow around a sail | 1986 | PE |
| | Coalescence of raindrops in an electrostatic field | 1999 | TPT |
| | A Life of Snow | 1969 | TPT |
| | Thermal patterns in the snow, Part I | 1976 | TPT |
| | The lightning discharge | 1976 | TPT |
| | Attraction or repulsion of a rod via electrical induction | 2011 | PE |
| | Fire resistance of framed buildings* | 2002 | PE |
| | Demonstration of a Faraday Cage Using a Metal Leaf Electroscope | 2019 | TPT |
| | Physics: clairvoyant of the \nEarth | 1991 | PE |
| | Thermal Patterns in the Snow: Icicles as Indicators of Heat Loss | 2008 | TPT |
| | The physics of snow and avalanche phenomena | 1980 | TPT |
| **Topic 10 - Circuits and electronics:** 2,4%"circuit" + 2,3%"voltage" + 2,0%"current" +1,3%"resistance" + 0,8%"output" + 0,7%"power" | | | |
| | Analogue-to-digital \nconversion made easy | 1992 | PE |
| | Investigating the active region of \ntransistor characteristics | 1993 | PE |
| | A second step to TTL circuitry | 1976 | PE |
| | Developing of a four-channel oscilloscope multiplexer for displaying digital sounds | 1991 | PE |
| | Microcomputer controlled experiments | 1982 | PE |
| | Arithmetic operations using TTL | 1976 | PE |
| | A microcomputer interface for external circuit control | 1983 | PE |
| | A simple voltage-to-frequency converter | 1974 | PE |
| | Output resistance | 1996 | TPT |
| | Cheap sensors for Pasco SW-500 interfaces | 2003 | PE |
| | Demonstrating electrical principles using an | 2001 | PE |
| | An elementary introduction to TTL | 1975 | PE |
| | Using a multimeter to study an RC circuit | 1995 | TPT |
| | Seeing rectifiers at work | 2002 | TPT |
| | Electronic counter‑timers for $6.00 per digit | 1977 | TPT |
| **Topic 11 - Thermodynamics:** 5,8%"temperature" + 4,5%"energy" + 2,4%"heat" + 1,1%"power" + 1,0%"cool" + 0,8%*"water" | | | |
| | Applications of refrigeration systems | 1986 | TPT |
| | Some energy problem problems and solutions | 1978 | TPT |
| | Power Generation: Nuclear, Oil, Coal | 2005 | TPT |
| | Removing the Mystery of Entropy and Thermodynamics - Part IV | 2012 | TPT |
| | The steam cycle used in electricity generation | 1967 | PE |
| | The physics of power stations: Part I: Fossil fuelled power stations | 1975 | PE |
| | Prospects for solar electricity | 1977 | TPT |
| | Removing the Mystery of Entropy and Thermodynamics - Part II | 2012 | TPT |
| | Graphic Representation of Quasi-Static Heat Exchange | 2018 | TPT |
| | The Law of Entropy Increase - A Lab Experiment | 2016 | TPT |
| | The second law of thermodynamics: heat engines and refrigerators | 1967 | PE |
| | A Conspectus on U.S. Energy | 2011 | TPT |
| | Desalting and Nuclear Energy | 1971 | TPT |
| | Commentary on "Figuring Physics/Rapid Evaporation" | 2007 | TPT |
| | Stirling engines for demonstration | 1982 | TPT |
| **Topic 12 - Astronomy:** 2,7%"earth" + 1,9%"sun" + 1,7%"star" + 1,1%"astronomy" + 1,0%"planet" + 1,0%"space" | | | |
| | Solar and lunar demonstrators | 2009 | PE |
| | Simulating the Phases of the Moon Shortly After Its Formation | 2014 | TPT |
| | An analemma experiment | 2000 | TPT |
| | Exoplanets | 2003 | PE |
| | Kepler's Third Law and NASA's Kepler Mission | 2015 | TPT |
| | Whither Does the Sun Rove? | 2011 | TPT |
| | A classroom activity and laboratory on astronomical scale | 2017 | TPT |
| | Diurnal Astronomy: Using Sticks and Threads to Find Our Latitude on Earth | 2012 | TPT |
| | Fixing the Shadows While Moving the Gnomon | 2015 | TPT |
| | The view from Mimas | 1988 | TPT |
| | Measuring global position using the Sun | 2014 | PE |
| | A stellar demonstrator | 2009 | PE |
| | Lookback time: Observing cosmic history | 1989 | TPT |
| | Dark energy | 2003 | PE |
| | Detecting Our Own Solar System from Afar | 2004 | TPT |
| **Topic 13 - Nuclear physics and medical physics**: 1,0%"radiation" + 0,8%"sample" + 0,7%"electron" + 0,7%"energy" + 0,7%"cell" + 0,7%"detector" | | | |
| | Semiconductor Detectors | 1970 | TPT |
| | Radiation physics and applications in therapeutic\nmedicine | 2001 | PE |
| | Fast Neutron Radiotherapy | 1973 | TPT |
| | Research in biophysics: The high resolution scanning electron microscope | 1975 | TPT |
| | Teaching about natural background radiation | 2013 | PE |
| | An illustrated guide to measuring radiocarbon from archaeological samples | 2004 | PE |
| | Chemical identification of surface monolayers | 1976 | TPT |
| | Radiotherapy by external beams of ionising radiation | 1978 | PE |
| | Radon adsorbed in activated charcoal—a simple and safe radiation source for teaching\npractical radioactivity in schools and colleges | 2012 | PE |
| | Radioactive consumer products in the classroom | 1995 | TPT |
| | Nuclear medicine | 2001 | PE |
| | Radiation exposure in our daily lives | 1977 | TPT |
| | Physics in nuclear medicine | 1978 | PE |
| | Diamond films and coatings: the flame deposition method and applications | 1997 | PE |
| | Molecular beam epitaxy | 1981 | PE |
| **Topic 14 - Fundamental particles and interactions:** 4,2%"particle" + 3,3%"energy" + 2,2%"electron" + 1,3%"mass" + 1,3%"quantum" + 1,1%"atom" | | | |
| | Particles, Feynman diagrams and all that | 2006 | PE |
| | Medical imaging | 1996 | PE |
| | The Standard Model | 1992 | PE |
| | A temporary organization of the subatomic particles | 1975 | TPT |
| | An article about a particle with charm? | 1975 | TPT |
| | Meet the family: the Leptons | 1997 | PE |
| | Teaching Elementary Particle Physics, PartII | 2011 | TPT |
| | Beta Decay | 1988 | TPT |
| | Looking for consistency in the construction and use of\nFeynman diagrams | 2001 | PE |
| | Exploring the standard model of particles | 2013 | PE |
| | Exploring quarks, gluons and the Higgs boson | 2013 | PE |
| | The origin of mass | 1996 | PE |
| | Assessment for learning in physics investigations: assessment criteria, questions and feedback in marking | 2006 | PE |
| | Let's have a coffee with the Standard Model of particle physics! | 2017 | PE |
| | The fundamental interactions of matter (for teachers) | 1977 | PE |
| **Topic 15 - History and Philosophy of Physics: Scientists, Books, and Ideas:** 0,7%"book" + 0,5%"world" + 0,5%"scientist" + 0,5%"theory" + 0,5%"scientific" + 0,4%"idea" | | | |
| | Resource Letter SL-1 on Science and Literature | 1966 | TPT |
| | Early science books and their women translators | 1998 | TPT |
| | Lady Newton' - an eighteenth century Marquise | 1996 | PE |
| | Albert Einstein: his life | 1979 | PE |
| | Where Credit is Due | 1972 | TPT |
| | Assimilation and Development in Arab Scientific Thought | 1966 | TPT |
| | Isaac Newton - 250th anniversary | 1977 | PE |
| | Einstein, relativity, and the press | 1980 | TPT |
| | Joseph Priestley, an early immigrant scientist | 1976 | TPT |
| | Physics and faith | 1987 | PE |
| | A non-believer looks at physics | 1987 | PE |
| | Where credit is due The Leaning Tower of Pisa Experiment | 1972 | TPT |
| | Benjamin Thompson, Count Rumford | 1976 | TPT |
| | A Copernicus Quiz | 1973 | TPT |
| | Galileo Galilei - Outline of His Life and Works | 1966 | TPT |
| **Topic 16 - Trends in Physics education: course structures and career pathways:** 0,8%"pupil" + 0,7%"subject" + 0,6%"education" + 0,6%"project" + 0,4%"examination" + 0,4%"unit" | | | |
| | TEC physics | 1980 | PE |
| | Examination results | 1969 | PE |
| | Occupation of successful candidates in the 1969 graduateship examination of The Institute of Physics | 1972 | PE |
| | Physics courses at Loughborough | 1978 | PE |
| | The Higher National Diploma in Applied Physics | 1969 | PE |
| | Joint-honours degrees at Manchester | 1978 | PE |
| | | 1972 | PE |
| | The Tyneside Physics Centre | 1967 | PE |
| | Physics degree courses | 1973 | PE |
| | Non-advanced further education | 1986 | PE |
| | The organization of applied physics degree courses in colleges of technology | 1966 | PE |
| | Physics in the polytechnics | 1980 | PE |
| | Technician engineers and technicians | 1972 | PE |
| | The examining of advanced level practical physics by the Joint Matriculation Board | 1972 | PE |
| | The CEI view | 1981 | PE |
| **Topic 17 - Programs, texts, and teaching strategies**: 1,1%"class" + 1,0%"program" + 1,0%"high_school" + 0,7%"college" + 0,7%"lab" + 0,7%"text" | | | |
| | NOTES: A Self-paced Program of Instruction at a two-year Institution | 1973 | TPT |
| | Old Physics Taught a New Way | 1971 | TPT |
| | The one‑year, non‑calculus physics course: An update* | 1988 | TPT |
| | Teaching physics: A human endeavor interview III | 1977 | TPT |
| | The Waterworld | 2004 | TPT |
| | Computer programs for physics lab | 1982 | TPT |
| | Nearly 1.4 Million High School Physics Students - Enrollments in AP and second-year courses up 26% even though number of graduates down in 2012-13 | 2014 | TPT |
| | "Retests": A better method of test corrections | 2011 | TPT |
| | Readability of college astronomy and physics texts III | 1983 | TPT |
| | Teaching physics: A human endeavor Interview IV | 1977 | TPT |
| | Update on the status of the one‑year, non‑calculus physics course | 1994 | TPT |
| | Student Selection of the Textbook for an Introductory Physics Course | 2007 | TPT |
| | Editorial Procedures | 2002 | TPT |
| | How do you select your physics textbook? | 1989 | TPT |
| | Teaching to teachers the course that they teach | 1985 | TPT |
| **Topic 18 - Demonstrations and apparatus in Physics teaching**: 3,0%"fig" + 1,2%"apparatus" + 0,8%"length" + 0,8%"demonstration" + 0,8%"cm" + 0,7%"spring" | | | |
| | Simple self‑lubricated electric motor for elementary physics lab | 1983 | TPT |
| | Rotational motion demonstrator | 1988 | TPT |
| | Three accessories for a rotating platform | 1980 | TPT |
| | The sand pendulum | 1987 | TPT |
| | Thermometer from a BIC ballpoint pen | 1977 | TPT |
| | NOTES: A Simple Camera-Panning Device | 1972 | TPT |
| | Apparatus for teaching physics. Classroom Foucault pendulum | 1996 | TPT |
| | Apparatus for Measuring Young's Modulus | 2003 | TPT |
| | Rotating stool mounted on a low-friction hub | 2000 | TPT |
| | The Iowa wave machines | 2010 | TPT |
| | Versatile celestial globe for introductory astronomy | 1980 | TPT |
| | Center of Mass of a Suspended Slinky: An Experiment | 2004 | TPT |
| | A standing wave simulator | 1989 | TPT |
| | Standing waves‑a new twist | 1983 | TPT |
| | Moment of inertia of a physical pendulum | 1996 | TPT |
| **Topic 19 - Data acquisition and measurement in Physics education**: 2,5%"data" + 1,7%"measurement" + 1,1%"graph" + 1,0%"sensor" + 0.,9%"model" + 0,7%"experimental" | | | |
| | A novel real-time data acquisition using an Excel spreadsheet in pendulum experiment tool with light-based timer | 2018 | PE |
| | Modeling human gait in high school | 2018 | PE |
| | Detecting position using ARKit | 2018 | PE |
| | Modeling Physics with Easy Java Simulations | 2007 | TPT |
| | An Arduino experiment to study free fall at schools | 2018 | PE |
| | Using | 2016 | PE |
| | A simple pendulum-based measurement of | 2018 | PE |
| | The Role of Data Range in Linear Regression | 2017 | TPT |
| | Deviation from the mean in teaching uncertainties | 2017 | PE |
| | Advanced tools for smartphone-based experiments: phyphox | 2018 | PE |
| | The Importance of Measurement Data Spacing | 2015 | TPT |
| | A new position sensor to analyze rolling motion using an iPhone | 2019 | PE |
| | Teaching physics using Microsoft Excel | 2017 | PE |
| | Driven damped harmonic oscillator resonance with an Arduino | 2017 | PE |
| | Creating interactive physics simulations using the power of GeoGebra | 2017 | TPT |
| **Topic 20 - Innovative learning strategies in Physics education:**1,8%"learn" + 1,0%"activity" + 1,0%"question" + 1,0%"concept" + 0,9%"group" + 0,9%"research" | | | |
| | Implications \nof research on learning for the education of\nprospective science and physics teachers | 2001 | PE |
| | Reflection on solutions in the form of refutation texts versus problem solving: the case of 8th graders studying simple electric circuits | 2017 | PE |
| | Using the Force Concept Inventory to monitor student\nlearning and to plan teaching | 2002 | PE |
| | Learning from Mistakes: The Effect of Students' Written Self-Diagnoses on Subsequent Problem Solving | 2016 | TPT |
| | Dealing with the Ambiguities of Science Inquiry | 2016 | TPT |
| | Physics textbooks: do they promote or inhibit students' creative thinking | 2015 | PE |
| | Self-diagnosis as a tool for supporting students' conceptual understanding and achievements in physics: the case of 8th-graders studying force and... | 2017 | PE |
| | Teaching renewable energy using online PBL in investigating its effect on behaviour towards energy conservation among Malaysian students: ANOVA re... | 2017 | PE |
| | Impact of Guided Reflection with Peers on the Development of Effective Problem Solving Strategies and Physics Learning | 2016 | TPT |
| | Best Practices for Administering Concept Inventories | 2017 | TPT |
| | Using categorization of problems as an instructional tool to help introductory students learn physics | 2016 | PE |
| | Periscope: Looking into Learning in Best-Practices Physics Classrooms | 2018 | TPT |
| | Preparing teachers to teach physics and physical science by\ninquiry | 2000 | PE |
| | Impact of interactive engagement on reducing the gender gap in quantum physics learning\noutcomes among senior secondary school students | 2012 | PE |
| | Large-scale studies on the transferability of general problem-solving skills and the pedagogic potential of physics | 2013 | PE |